\documentclass[referee,a4paper,12pt,traditabstract]{swsc} 

\usepackage{graphicx}
\usepackage{txfonts}
\usepackage{subfigure}
\usepackage{epstopdf}
\usepackage[displaymath,mathlines]{lineno}
\usepackage[authoryear,round]{natbib}
\usepackage[backref]{hyperref}
\usepackage{url}
\usepackage{gensymb}
\usepackage{footmisc}

\hypersetup{colorlinks=true,citecolor=cyan,urlcolor=cyan,linkcolor=blue}

\begin{document}


\title{A LOFAR Observation of Ionospheric Scintillation from Two Simultaneous Travelling Ionospheric Disturbances}

\titlerunning{LOFAR Ionospheric Scintillation}

\authorrunning{Fallows et al.}

\author{
    R.A. Fallows\inst{1}\footnote{Corresponding author: \href{mailto:fallows@astron.nl}{fallows@astron.nl}}
    \and
    B. Forte\inst{2}
    \and
    I. Astin\inst{2}
    \and
    T. Allbrook\inst{2}\footnote{Now at BAE Systems (operation) Ltd.}
    \and
    A. Arnold\inst{2}\footnote{Now an independent researcher}
    \and
    A. Wood\inst{3}
    \and
    G. Dorrian\inst{4}
    \and
    M. Mevius\inst{1}
    \and
    H. Rothkaehl\inst{5}
    \and
    B. Matyjasiak\inst{5}
    \and
    A. Krankowski\inst{6}
    \and
    J.M. Anderson\inst{7,8}
    \and
    A. Asgekar\inst{9}
    \and
    I.M. Avruch\inst{10}
    \and
    M.J.Bentum\inst{1}
    \and
    M.M. Bisi\inst{11}
    \and
    H.R. Butcher\inst{12}
    \and
    B. Ciardi\inst{13}
    \and
    B. Dabrowski\inst{6}
    \and
    S. Damstra\inst{1}
    \and
    F. de Gasperin\inst{14}
    \and
    S. Duscha\inst{1}
    \and
    J. Eisl\"offel\inst{15}
    \and
    T.M.O Franzen\inst{1}
    \and
    M.A. Garrett\inst{16,17}
    \and
    J.-M. Grie{\ss}meier\inst{18,19}
    \and
    A.W. Gunst\inst{1}
    \and
    M. Hoeft\inst{15}
    \and
    J.R.\ H\"orandel\inst{20,21,22}
    \and
    M. Iacobelli\inst{1}
    \and
    H.T. Intema\inst{17}
    \and
    L.V.E. Koopmans\inst{23}
    \and
    P. Maat\inst{1}
    \and
    G. Mann\inst{24}
    \and
    A. Nelles\inst{25,26}
    \and
    H. Paas\inst{27}
    \and
    V.N. Pandey\inst{1,23}
    \and
    W. Reich\inst{28}
    \and
    A. Rowlinson\inst{1,29}
    \and
    M. Ruiter\inst{1}
    \and
    D.J. Schwarz\inst{30}
    \and
    M. Serylak\inst{31,32}
    \and
    A. Shulevski\inst{29}
    \and
    O.M. Smirnov\inst{33,31}
    \and
    M. Soida\inst{34}
    \and
    M. Steinmetz\inst{24}
    \and
    S. Thoudam\inst{35}
    \and
    M.C. Toribio\inst{36}
    \and
    A. van Ardenne\inst{1}
    \and
    I.M. van Bemmel\inst{37}
    \and
    M.H.D. van der Wiel\inst{1}
    \and
    M.P. van Haarlem\inst{1}
    \and
    R.C. Vermeulen\inst{1}
    \and
    C. Vocks\inst{24}
    \and
    R.A.M.J. Wijers\inst{29}
    \and
    O. Wucknitz\inst{28}
    \and
    P. Zarka\inst{38}
    \and
    P. Zucca\inst{1}
    }

\institute{
    ASTRON - the Netherlands Institute for Radio Astronomy, Oude Hoogeveensedijk 4, 7991 PD Dwingeloo, the Netherlands 
    \and
    Department of Electronic and Electrical Engineering, University of Bath, Claverton Down, Bath, BA2 7AY, UK
    \and
    School of Science and Technology, Nottingham Trent University, Clifton Lane, Nottingham, NG11 8NS, UK
    \and
    Space Environment and Radio Engineering, School of Engineering, The University of \\ Birmingham, Edgbaston, Birmingham, B15 2TT, UK
    \and
    Space Research Centre, Polish Academy of Sciences, Bartycka 18A, 00-716 Warsaw, Poland
    \and
    Space Radio-Diagnostics Research Centre, University of Warmia and Mazury, ul. Romana Prawocheskiego 9, 10-719 Olsztyn, Poland
    \and
    Technische Universit\"{a}t Berlin, Institut f\"{u}r Geod\"{a}sie und Geoinformationstechnik, Fakult\"{a}t VI, Sekr. H 12, Hauptgeb\"{a}ude Raum H 5121, Stra{\ss}e des 17. Juni 135, 10623 Berlin, Germany
    \and
    GFZ German Research Centre for Geosciences, Telegrafenberg, 14473 Potsdam, Germany
    \and
    Shell Technology Center, Bangalore, India
    \and
    Science and Technology B.V., Delft, the Netherlands
    \and
    RAL Space, UKRI STFC, Rutherford Appleton Laboratory, Harwell Campus, Oxfordshire, OX11 0QX, UK
    \newpage
    \and
    Mt Stromlo Observatory, Research School of Astronomy and Astrophysics, Australian National University, Cotter Road, Weston Creek, ACT 2611, Australia
    \and
    Max Planck Institute for Astrophysics, Karl-Schwarzschild-Str. 1, 85748 Garching, Germany
    \and
    Hamburger Sternwarte, Universit\"{a}t Hamburg, Gojenbergsweg 112, 21029, Hamburg, Germany
    \and
    Th\"uringer Landessternwarte, Sternwarte 4, D-07778 Tautenburg, Germany
    \and
    Jodrell Bank Centre for Astrophysics (JBCA), Department of Physics \& Astronomy, Alan Turing Building, Oxford Road, University of Manchester, Manchester M139PL, UK
    \and
    Leiden Observatory, Leiden University, PO Box 9513, NL-2300 RA Leiden, The Netherlands
    \and
    LPC2E - Universit\'{e} d'Orl\'{e}ans /  CNRS, 45071 Orl\'{e}ans cedex 2, France
    \and
    Station de Radioastronomie de Nan\c{c}ay, Observatoire de Paris, PSL Research University, CNRS, Univ. Orl\'{e}ans, OSUC, 18330 Nan\c{c}ay, France
    \and
    Radboud University, Department of Astrophysics/IMAPP, P.O. Box 9010, 6500 GL Nijmegen, The Netherlands
    \and
    Nikhef, Science Park 105, 1098 XG Amsterdam, The Netherlands
    \and
    Vrije Universiteit Brussel, Astronomy and Astrophysics Research Group, Pleinlaan 2,  1050 Brussel, Belgium
    \and
    Kapteyn Astronomical Institute, University of Groningen, P.O.Box 800, 9700AV Groningen, the Netherlands
    \and
    Leibniz-Institut f\"{u}r Astrophysik Potsdam, An der Sternwarte 16, D-14482 Potsdam, Germany
    \and
    ECAP, Friedrich-Alexander-Universit\"{a}t Erlangen-N\"{u}rnberg, Erwin-Rommel-Str. 1, \\ 91054 Erlangen, Germany
    \and
    DESY, Platanenallee 6, 15738 Zeuthen, Germany
    \and
    CIT, Rijksuniversiteit Groningen,  Nettelbosje 1, 9747 AJ Groningen, The Netherlands
    \and
    Max-Planck-Institut f\"{u}r Radioastronomie, Auf dem H\"{u}gel 69, 53121 Bonn, Germany
    \and
    Anton Pannekoek Institute, University of Amsterdam, Postbus 94249, 1090 GE Amsterdam, The Netherlands
    \and
    Fakult\"{a}t f\"{u}r Physik, Universit\"{a}t Bielefeld, Postfach 100131, 33501 Bielefeld, Germany
    \and
    South African Radio Astronomy Observatory, 2 Fir Street, Black River Park, Observatory, Cape Town, 7925, South Africa
    \and
    Department of Physics and Astronomy, University of the Western Cape, Cape Town 7535, South Africa
    \and
    Department of Physics and Electronics, Rhodes University, PO Box 94, Makhanda, 6140, South Africa
    \and
    Jagiellonian University in Krak\'{o}w, Astronomical Observatory, ul. Orla 171, PL 30-244 Krak\'{o}w, Poland
    \and
    Department of Physics, Khalifa University, PO Box 127788, Abu Dhabi, United Arab Emirates
    \and
    Department of Space, Earth and Environment, Chalmers University of Technology, Onsala Space Observatory, SE-439 92 Onsala, Sweden
    \and
    JIVE, Joint Institute for VLBI-ERIC, Oude Hoogeveensedijk 4, 7991 PD Dwingeloo, the Netherlands
    \and
    LESIA \& USN, Observatoire de Paris, CNRS, PSL, SU/UP/UO, 92195 Meudon, France
    }

\date{Received November 30, 2019}

\abstract{
    This paper presents the results from one of the first observations of ionospheric scintillation taken using the Low-Frequency Array (LOFAR).  The observation was of the strong natural radio source Cassiopeia A, taken overnight on 18-19 August 2013, and exhibited moderately strong scattering effects in dynamic spectra of intensity received across an observing bandwidth of 10-80\,MHz.  Delay-Doppler spectra (the 2-D FFT of the dynamic spectrum) from the first hour of observation showed two discrete parabolic arcs, one with a steep curvature and the other shallow, which can be used to provide estimates of the distance to, and velocity of, the scattering plasma.  A cross-correlation analysis of data received by the dense array of stations in the LOFAR ``core'' reveals two different velocities in the scintillation pattern: a primary velocity of $\sim$20-40\,m\,s$^{-1}$ with a north-west to south-east direction, associated with the steep parabolic arc and a scattering altitude in the F-region or higher, and a secondary velocity of $\sim$110\,m\,s$^{-1}$ with a north-east to south-west direction, associated with the shallow arc and a scattering altitude in the D-region.  Geomagnetic activity was low in the mid-latitudes at the time, but a weak sub-storm at high latitudes reached its peak at the start of the observation.  An analysis of Global Navigation Satellite Systems (GNSS) and ionosonde data from the time reveals a larger--scale travelling ionospheric disturbance (TID), possibly the result of the high--latitude activity, travelling in the north-west to south-east direction, and, simultaneously, a smaller--scale TID travelling in a north-east to south-west direction, which could be associated with atmospheric gravity wave activity.  The LOFAR observation shows scattering from both TIDs, at different altitudes and propagating in different directions.  To the best of our knowledge this is the first time that such a phenomenon has been reported.   
  }
   

\maketitle


\section{Introduction}

    Radio waves from compact sources can be strongly affected by any ionised medium through which they pass.  Refraction through large-scale density structures in the medium leads to strong lensing effects where the radio source appears, if imaged, to focus, de-focus and change shape as the density structures in the line of sight themselves move and change.  Diffraction of the wavefront by small-scale density structures leads to variations building up in the intensity of the wavefront with distance from the scattering medium, due to interference between the scattered waves, an effect known as scintillation.  Observations of all these effects thus contain a great deal of information on the medium through which the radio waves have passed, including the large-scale density, turbulence, and the movement of the medium across the line of sight.  Since the second world war, a large number of studies have shown the effect of ionospheric density variations on radio signals, as reviewed by \citet{Aarons:1982}, and this can lead to disruption for applications using Global Navigation Satellite Systems (GNSS, e.g., GPS), as thoroughly reviewed by, e.g., \citet{Hapgood:2017}.  The Low-Frequency Array (LOFAR - \citet{vanHaarlemetal:2013}) is Europe's largest low-frequency radio telescope, operating across the frequency band 10--250\,MHz, and with a dense array of stations in the Netherlands and, at the time of writing, 13 stations internationally from Ireland to Poland.  It was conceived and designed for radio astronomy but, at these frequencies, the ionosphere can also have a strong effect on the radio astronomy measurement \citep{deGasperinetal:2018}.  Ionospheric scintillation, which is rarely seen over the mid-latitudes on the high-frequency signals of GNSS, is seen almost continually in observations of strong natural radio sources by LOFAR.
    
    The wide bandwidth available with LOFAR enables an easy and direct assessment of scattering conditions and how they change in a given observation, including whether scattering is weak or strong, or refractive effects dominate, and enables further information to be gleaned from delay-Doppler spectra (the 2-D FFT of a dynamic spectrum, termed variously as the ``scattering function'', ``generalised power spectrum'', or ``secondary spectrum'' - here we use the term ``delay-Doppler'' spectrum as this clearly describes what the spectrum shows).  In observations of interstellar scintillation these spectra can exhibit discrete parabolic arcs which can be modelled to give information on the distance to the scattering ``screen'' giving rise to the scintillation and its velocity across the line of sight \citep{Stinebringetal:2001, Cordesetal:2006}.  Broadband observations of ionospheric scintillation are not common, but such arcs have been observed using the Kilpisj\"arvi Atmospheric Imaging Receiver Array (KAIRA, \citet{KAIRA-reference-paper:2014} -- an independent station built using LOFAR hardware in arctic Finland) in a study by \citet{Fallowsetal:2014}.  Model spectra produced by \citet{KneppNickisch:2009} have also illustrated parabolic arc structures, particularly in the case of scattering from a thin scattering screen. 
    
    The wide spatial distribution of LOFAR stations also enables scintillation conditions at these observing frequencies to be sampled over a large part of western Europe.  A dense ``core'' of 24 stations, situated near Exloo in the north-east of the Netherlands, over an area with a diameter of $\sim$3.5\,km further provides a more detailed spatial view of the scintillation pattern in its field of view. 
    
    LOFAR thus enables detailed studies of ionospheric scintillation to be undertaken which can both reveal details which would be unavailable to discrete-frequency observations such as those taken using GNSS receivers, and act as a low-frequency complement to these observations to probe potentially different scattering scales.
    
    A number of different phenomena can lead to scattering effects in radio wave propagation through the mid-latitude ionosphere: Ionisation structures due to gradients in the spatial distribution of the plasma density can arise from a southward expansion of the auroral oval or from large- to small- scale travelling ionospheric disturbances (TIDs).  Large-scale TIDs (LSTIDs) with wavelengths of about 200\,km typically propagate southward after forming in the high-latitude ionosphere in response to magnetic disturbances (e.g. storms or sub-storms, \citet{Tsugawaetal:2004}). On the other hand, medium-scale TIDs (MSTIDs) seem to form in response to phenomena occurring in the neutral atmosphere triggering atmospheric gravity waves (AGWs), which then propagate upwards to generate TIDs at ionospheric heights \citep{Kelley:2009}. The morphology of MSTIDs varies with local time, season, and magnetic longitude.  Their propagation shows irregular patterns that vary on a case-by-case basis, although they commonly seem to propagate mainly equatorward during winter daytime and westward during summer night-time \citep{HernandezPajaresetal:2006, HernandezPajaresetal:2012, Tsugawaetal:2007, SaitoFukao:1998, Emardsonetal:2013}. Smaller-scale ionisation gradients, likely associated with the Perkins instability \citep{Kelley:2009, Kelley:2011}, can then form as a consequence of the presence of MSTIDs, potentially leading to scintillation at LOFAR frequencies.

    In this paper, we perform an in-depth analysis of ionospheric scintillation seen in an observation of the strong natural radio source Cassiopeia A (Cas A) overnight on 18-19 August 2013.  This observation was amongst the first of its kind taken with LOFAR and exhibited quite strong scattering effects across the 10-80\,MHz band.  The purpose of this paper is both technical and scientific:  We first describe the observation itself, and then demonstrate several techniques to analyse LOFAR data and show how these can bring out the details of ionospheric structures.  Finally, we use supporting data from GNSS and ionosondes to get a broader picture of conditions in the ionosphere at the time and how these give rise to the scintillation seen by LOFAR.

\section{The LOFAR Observation}

    LOFAR observed Cas A (Right Ascension 23h23m24s, Declination +58\degree48'54") between 21:05\,UT on 18 August 2013 and 04:05\,UT on 19 August 2013, recording dynamic spectra from each individual station with a sampling time of 0.083\,s over the band 2.24-97.55\,MHz from each available station.  The observing band was sampled with 7808 channels of 12.207\,kHz each, but averaged over each successive 16-channel block to 488 subbands of 195.3125\,kHz for the analyses described in this paper.  At the time of observation the available stations were the 24 stations of the LOFAR ``core'', 13 ``remote'' stations across the north-east of the Netherlands, and the international stations at Effelsburg, Unterweilenbach, Tautenburg, Potsdam, and J\"ulich (Germany), Nan\c{c}ay (France), Onsala (Sweden), and Chilbolton (UK).  The reader is referred to \citet{vanHaarlemetal:2013} for full details of the LOFAR receiving system.  The raw data for this observation can be obtained from the LOFAR long-term archive (\url{lta.lofar.eu}); observation ID L169059 under project ``IPS''.
    
\begin{figure}
        \centering
        \subfigure[]{\label{fig:timeseries}\includegraphics[width=.9\linewidth]{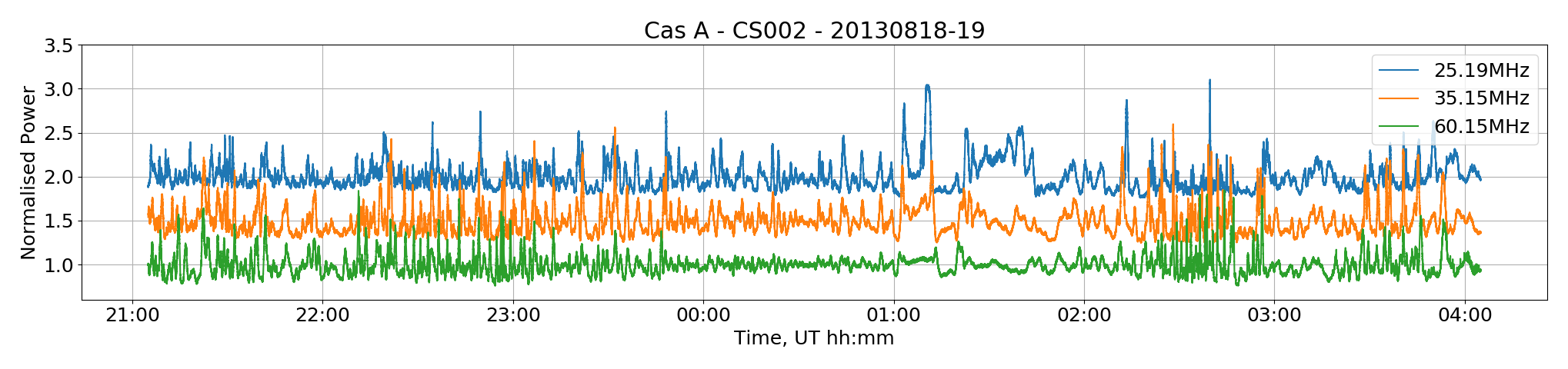}}
        \\
        \subfigure[]{\label{fig:pspec1}\includegraphics[width=.45\linewidth]{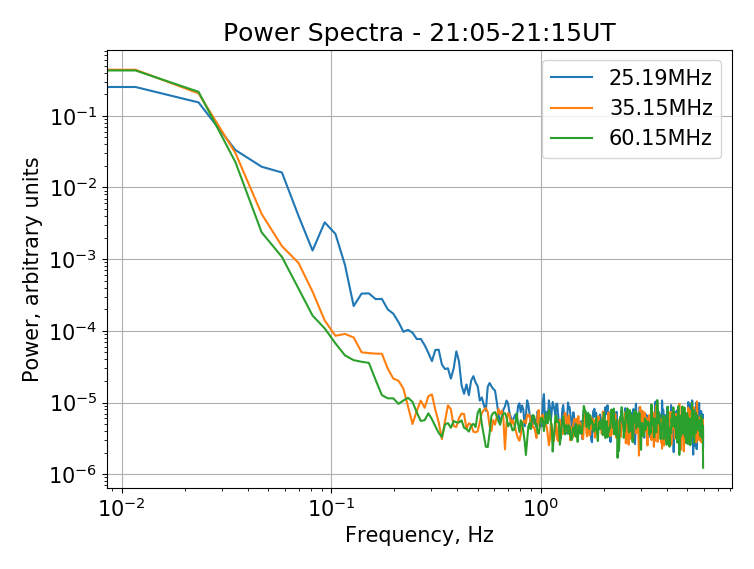}}
        \subfigure[]{\label{fig:pspec2}\includegraphics[width=.45\linewidth]{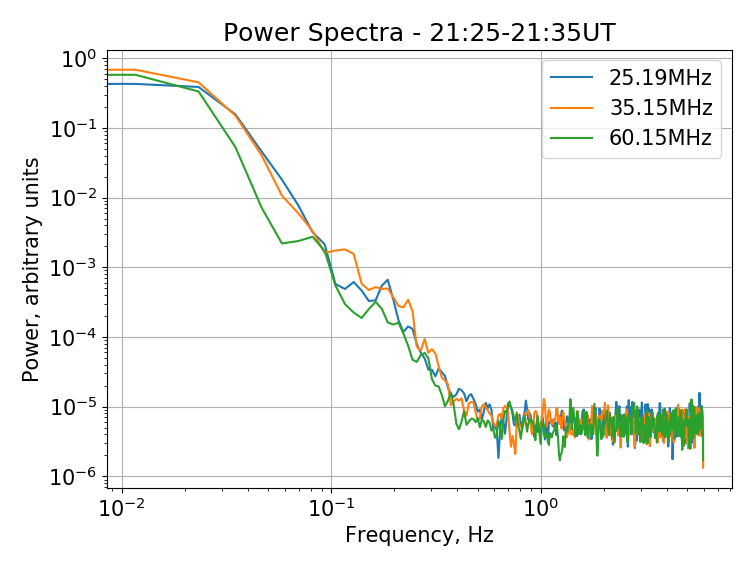}}
        \caption{\subref{fig:timeseries} Time series of intensity received at three discrete frequencies by LOFAR station CS002 during the observation of Cas A on 18-19 August 2013, plus, \subref{fig:pspec1} and \subref{fig:pspec2}, power spectra of two 10-minute periods within these time series'.}
        \label{fig:timespectra}
    \end{figure}

    We first illustrate the data in a more traditional sense.  Figure \ref{fig:timespectra} shows time series' at three discrete observing frequencies of the data taken by LOFAR station CS002, at the centre of the core, and their associated power spectra.  The power spectra show a fairly typical shape for intensity scintillation:  An initial flat section at the lowest spectral frequencies represents scattering from larger-scale density structures which are close enough to the observer that the scattered waves have not had the space to fully interfere to develop a full intensity scintillation pattern; the turnover (often termed the ``Fresnel Knee'') indicates the largest density scales for which the intensity scintillation pattern has fully formed; this is followed by a power-law in the spectra illustrating the cascade from larger to smaller density scales, which is cut off in these spectra by white noise due to the receiving system (the flat section covering high spectral frequencies).
    
\begin{figure}
        \centering
        \subfigure[]{\label{fig:fullspectrum}\includegraphics[width=.9\linewidth]{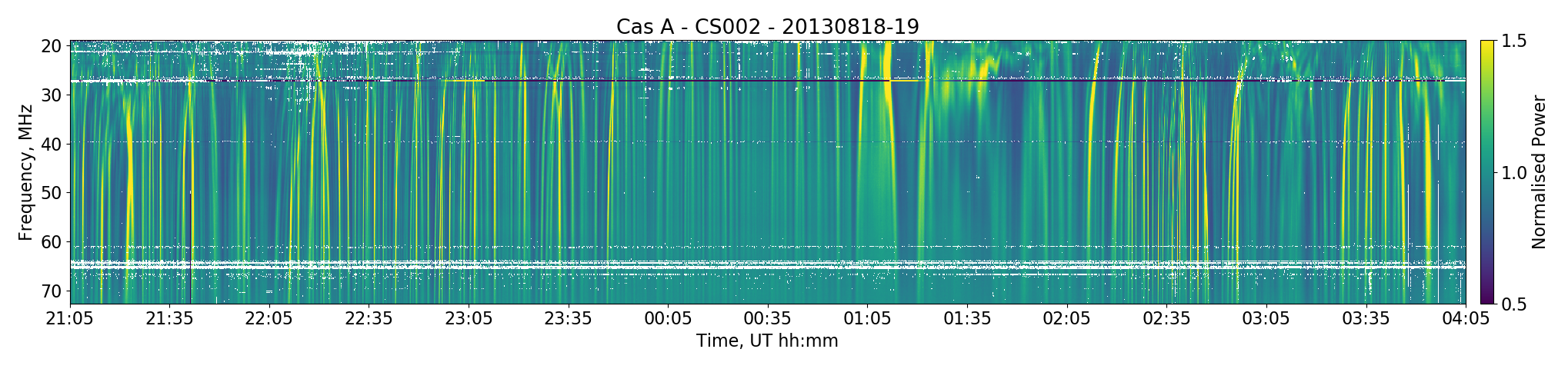}}
        \\
        \subfigure[]{\label{fig:dynspec2122}\includegraphics[width=.9\linewidth]{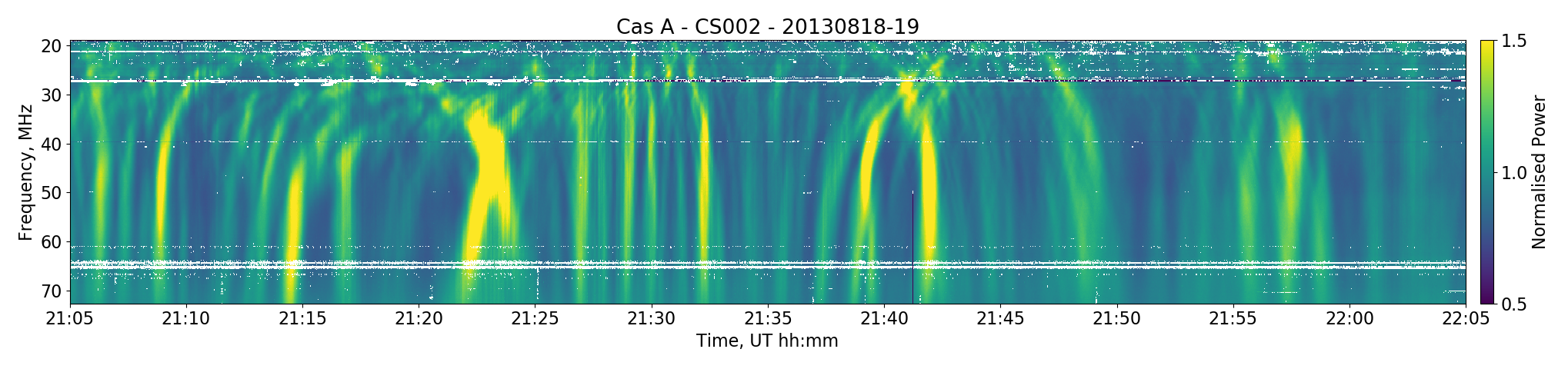}}
        \\
        \subfigure[]{\label{fig:dynspec0102}\includegraphics[width=.9\linewidth]{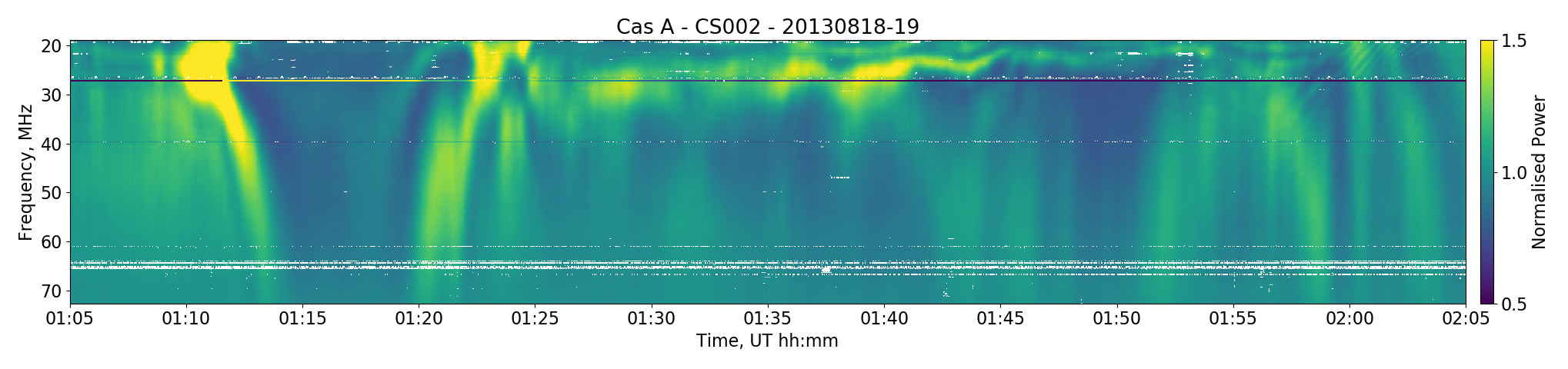}}
        \\
        \subfigure[]{\label{fig:dynspec02300330}\includegraphics[width=.9\linewidth]{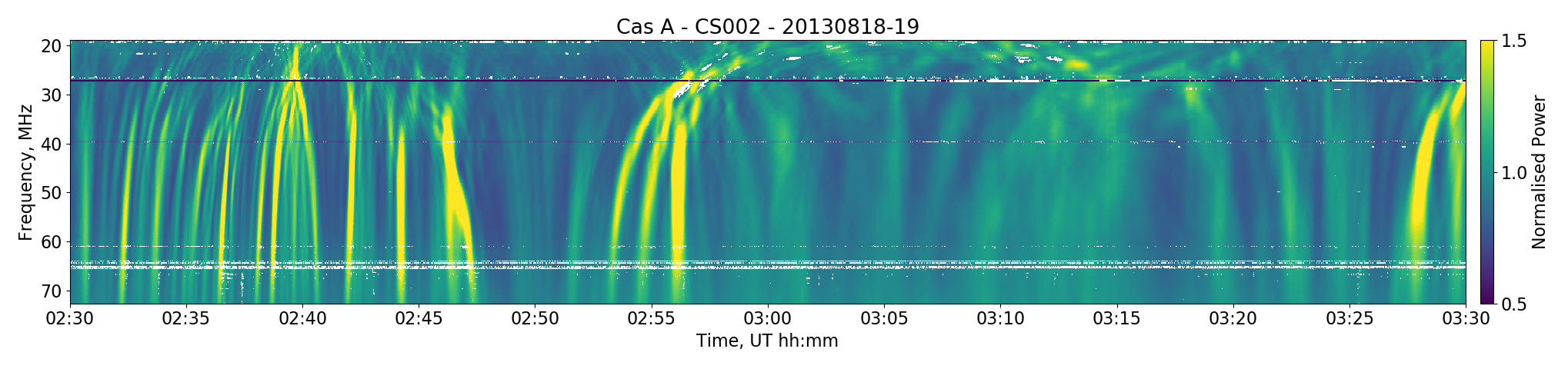}}
        \caption{Dynamic spectra of normalised intensity data taken by LOFAR station CS002 during the observation of Cas A on 18-19 August 2013.  The dynamic spectrum of the entire observing period is given at the top, with zooms into three different hours of observation below to illustrate the range of conditions seen.  White areas within the plots indicate where RFI was identified.}
        \label{fig:dynspecs}
    \end{figure}

    However, the advantage of observing a natural radio source with LOFAR is that full dynamic spectra can be produced covering the full observed band.  Dynamic spectra of data taken by LOFAR station CS002 are presented in Figure \ref{fig:dynspecs}, which includes a dynamic spectrum of the full observation, alongside more detailed views of three different single hours of the observation to illustrate the range of scattering conditions seen.  The strength of the scattering can be seen much more clearly in this view, compared to time series' from discrete observing frequencies.  In general, scattering appears weak in this observation at the highest observing frequencies (where intensity remains highly correlated across the observing band) with a transition to strong scattering conditions as the observing frequency decreases.  The frequency range displayed in these dynamic spectra is restricted to exclude the radio--frequency interference (RFI) which dominates below about 20\,MHz and a fade in signal strength at the higher frequencies due to the imposition of a hard filter to exclude the FM waveband.

    RFI is still visible as white areas within the plots.  These were identified by applying a median filter to the data using a window of (19.5\,MHz $\times$ 4.2\,s) to flatten out the scintillation pattern and then applying a threshold to identify the RFI.  This method appears to be quite successful at identifying the RFI without also falsely identifying strong peaks in the scintillation as RFI.  For subsequent analysis the RFI data points are replaced by an interpolation from nearby data, using the Python Astropy \citep{astropy:2013, astropy:2018} library routine, ``interpolate\_replace\_nans''.  Normalisation of the data, to correct for long-period temporal variations in the system (e.g., gain variations resulting from the varying sensitivity of the receiving antenna array with source elevation), is carried out after RFI excision by dividing the intensities for each single frequency subband by a fitted 3$^{rd}$-order polynomial.  
    
    When analysing the data, a variety of scattering conditions are observed during the course of the observation, as indicated in Figure \ref{fig:dynspecs}.  Different conditions also naturally occurred over the various international stations compared to those observed over the Dutch part of LOFAR.  In this paper we therefore focus our analysis on only the first hour of observation and the measurements taken by the 24 core stations. This allows us to demonstrate the analysis techniques and to investigate the reason for the scintillation seen over this interval.  Observations from later in this dataset undoubtedly show other effects and may be discussed in a subsequent publication.  
    
\section{LOFAR Data Analysis Methods and Results}

    \subsection{Delay-Doppler Spectra}
    
\begin{figure}
            \centering
            \subfigure[]{\label{fig:fullspectrum14}\includegraphics[width=.32\linewidth]{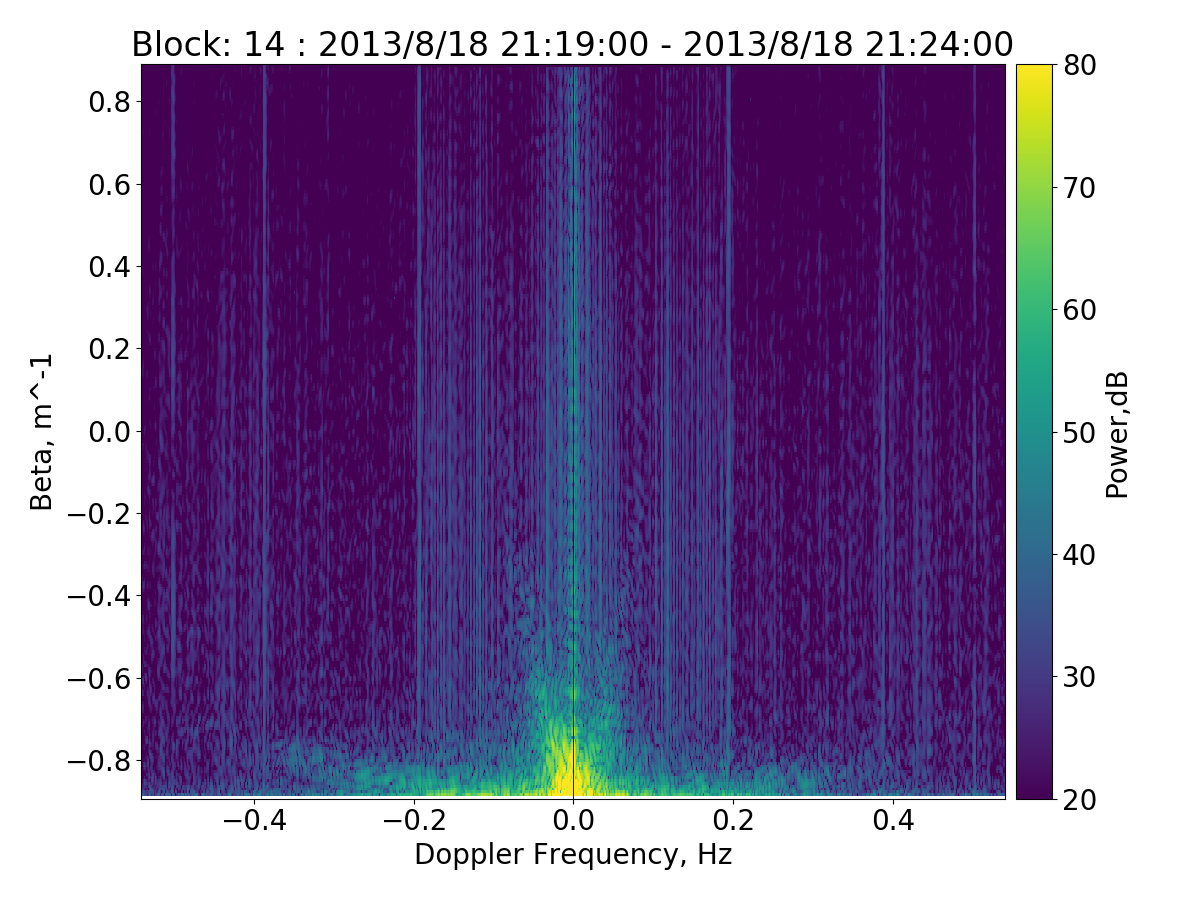}}
            \subfigure[]{\label{fig:fullspectrum21}\includegraphics[width=.32\linewidth]{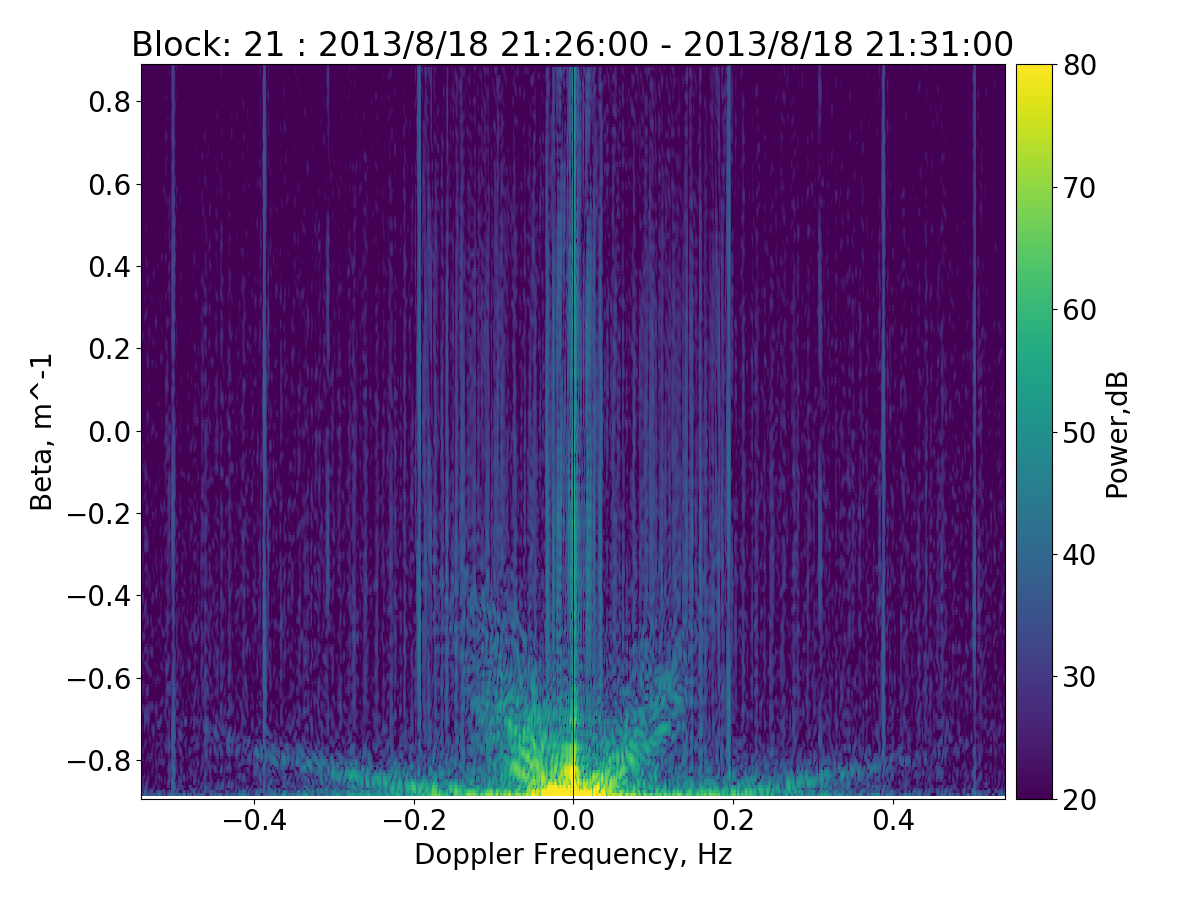}}
            \subfigure[]{\label{fig:fullspectrum30}\includegraphics[width=.32\linewidth]{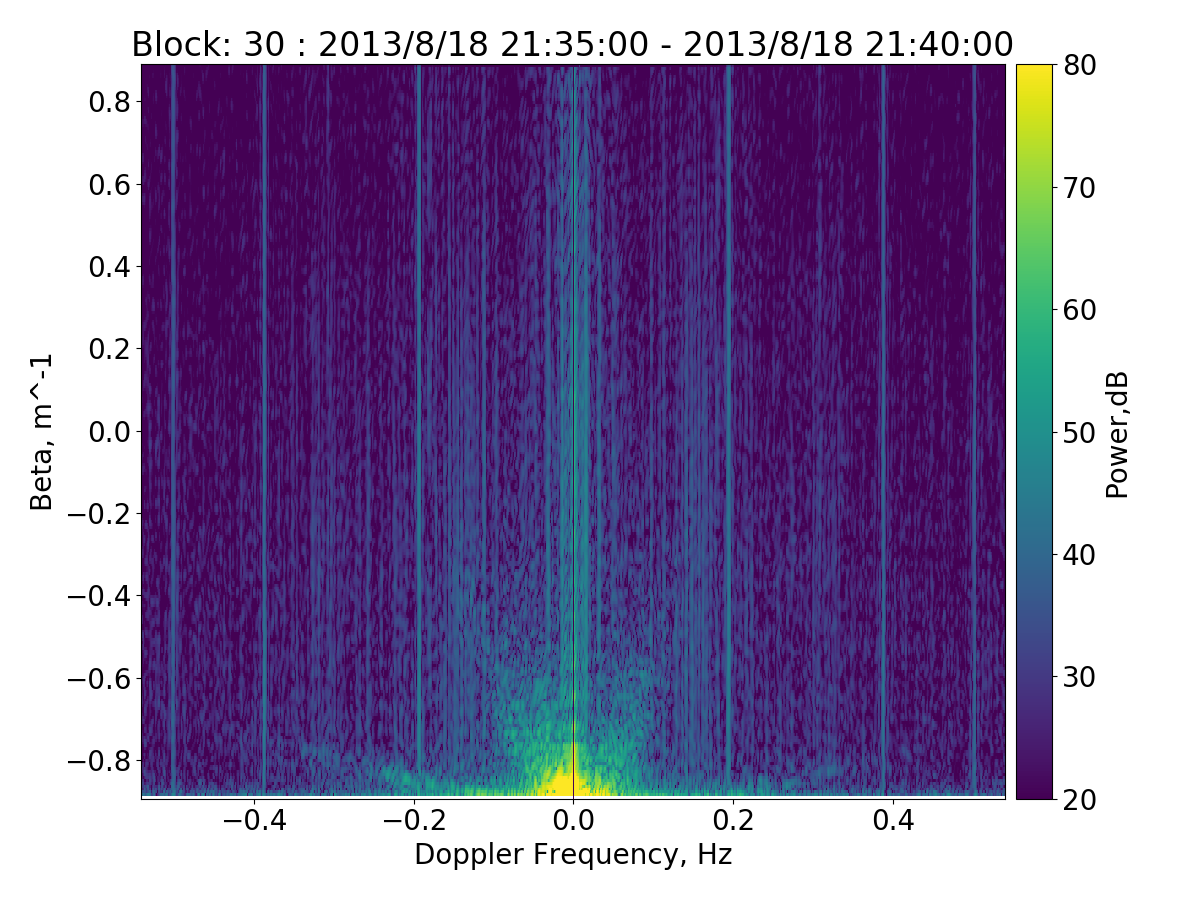}}
            \caption{Example delay-Doppler spectra from the first hour of observation, taken using five-minute chunks of the dynamic spectrum from CS103 over the frequency band 28.5-64.1\,MHz.}
            \label{fig:delaydoppler}
        \end{figure}

        The first stage of analysis was the calculation of delay--Doppler spectra: These were created from the dynamic spectra using five-minute time slices, advancing every minute through the observation, following the methods described in \citet{Fallowsetal:2014}.  To avoid regions more heavily contaminated by RFI, the frequency band used was restricted to 28.5--64.1\,MHz.  Example spectra from the first hour are presented in Figure \ref{fig:delaydoppler}.
        
        The spectra show two clear arcs: the first is a steeper arc which varies in curvature throughout the first hour (henceforth labelled for convenience as the ``primary arc''); the second is a very shallow arc (henceforth labelled as the ``secondary arc'') which remains stable for the first 40 minutes of the observation before fading away.  By the end of the first hour of observation the primary arc also becomes less distinctive for a short while before the delay--Doppler spectra again show distinctive structure, including a return of the secondary arc.  
        
\begin{figure}
            \centering
            \includegraphics[width=.9\linewidth]{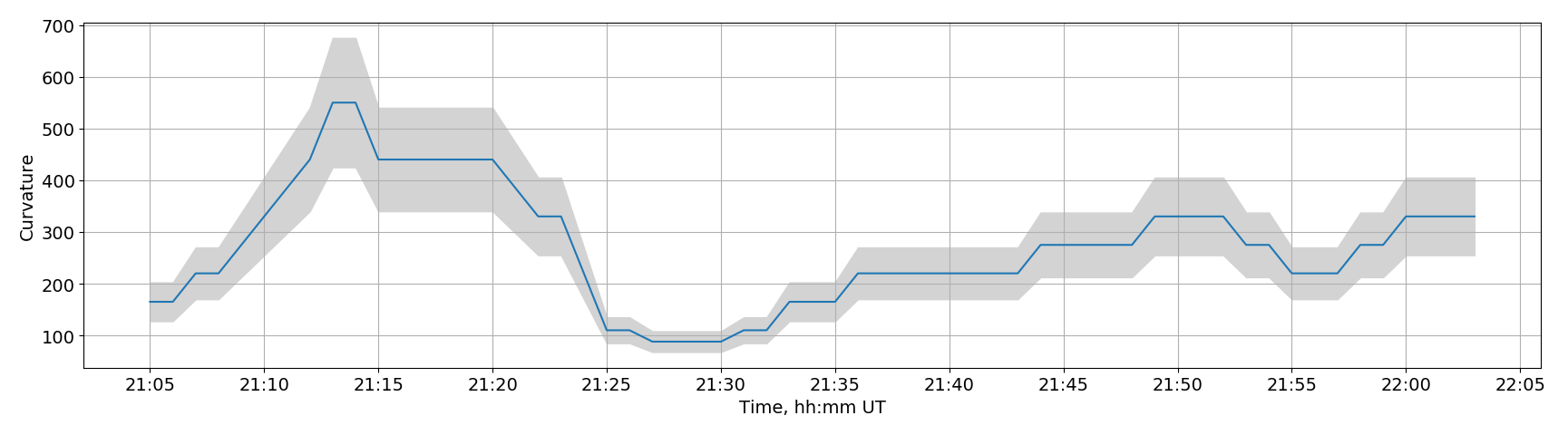}
            \caption{Curvatures of the steeper arc seen in delay-Doppler spectra calculated using data from CS103, from simple parabolas fitted by eye.  The grey bounds represent an estimated error.}
            \label{fig:curves}
        \end{figure}

        The variability of the curvature of the primary arc appears to follow a wave--like pattern during this part of the observation, as displayed in Figure \ref{fig:curves}.  Here, simple parabolas involving only the square term ($y = Cx^2$ where $C$ is the curvature) were plotted with various curvatures until a reasonable eyeball fit was achieved, and the resulting curvatures plotted for every minute of observation for the first hour.  It proved impossible to achieve reasonable fits using least-squares methods due to confusion from non--arc structure in the spectra: Fitting curvatures to these scintillation arcs is a well--known problem in the interstellar scintillation field and new methods of attempting this were presented at a recent workshop, but they are not easily described and have yet to be published.  Hence, we do not attempt their application here.
        
        The presence of two scintillation arcs likely indicates that scattering is dominated by two distinct layers in the ionosphere.  A simple analysis, as described in \citet{Fallowsetal:2014}, can be used to estimate the altitude of the scattering region with a basic formula relating arc curvature $C$ to velocity $V$ and distance $L$ along the line of sight to the scattering region \citep{Cordesetal:2006}: 
        
        \begin{equation}
            \label{eqn:scintarc}
            L = 2 C V^{2}
        \end{equation}
        The square term for the velocity illustrates the importance of gaining a good estimate of velocity to be able to accurately estimate the altitude of the scattering region via this method.
        
    \subsection{Scintillation Pattern Flow}
        
\begin{figure}
            \centering
            \subfigure{\includegraphics[width=.266\linewidth]{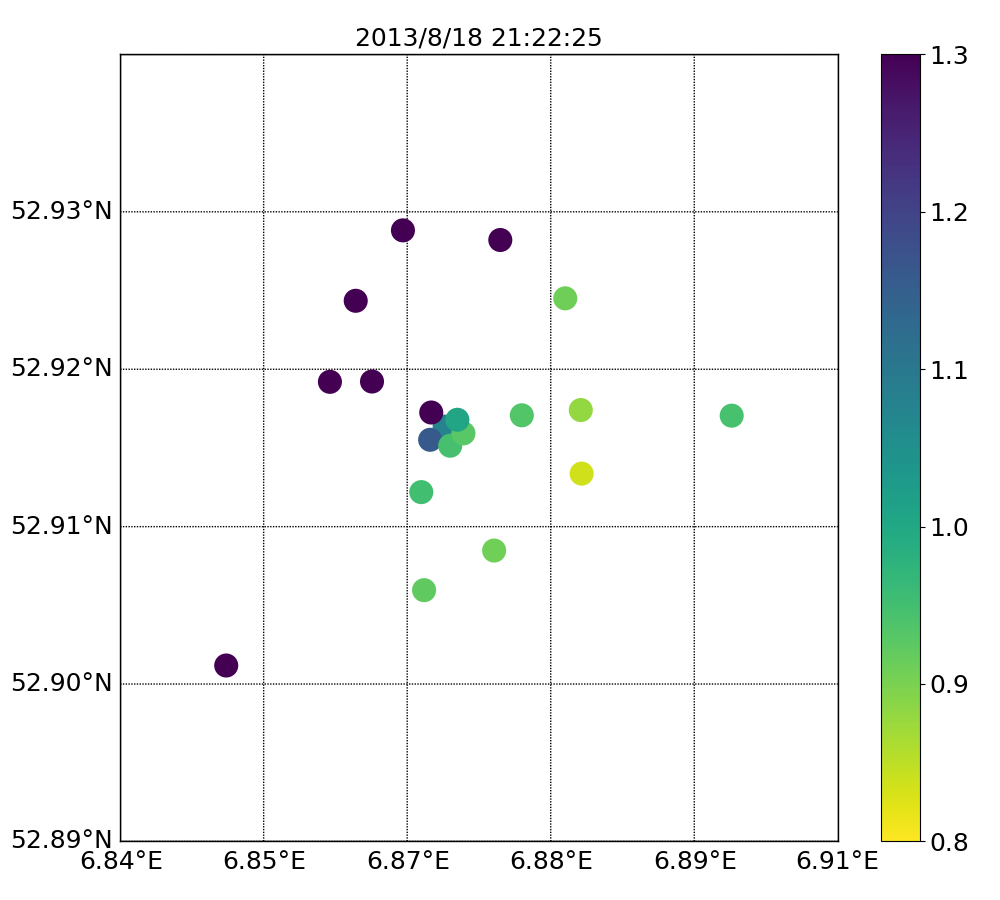}}
            \subfigure{\includegraphics[width=.266\linewidth]{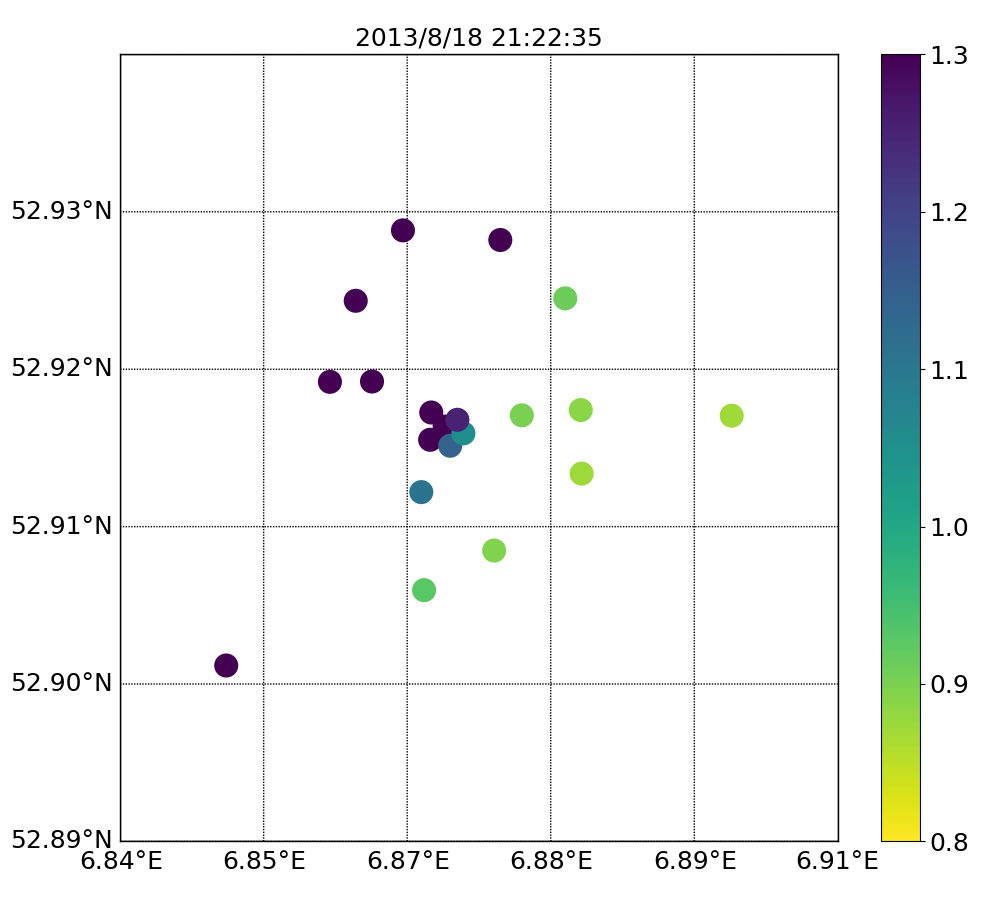}}
            \subfigure{\includegraphics[width=.266\linewidth]{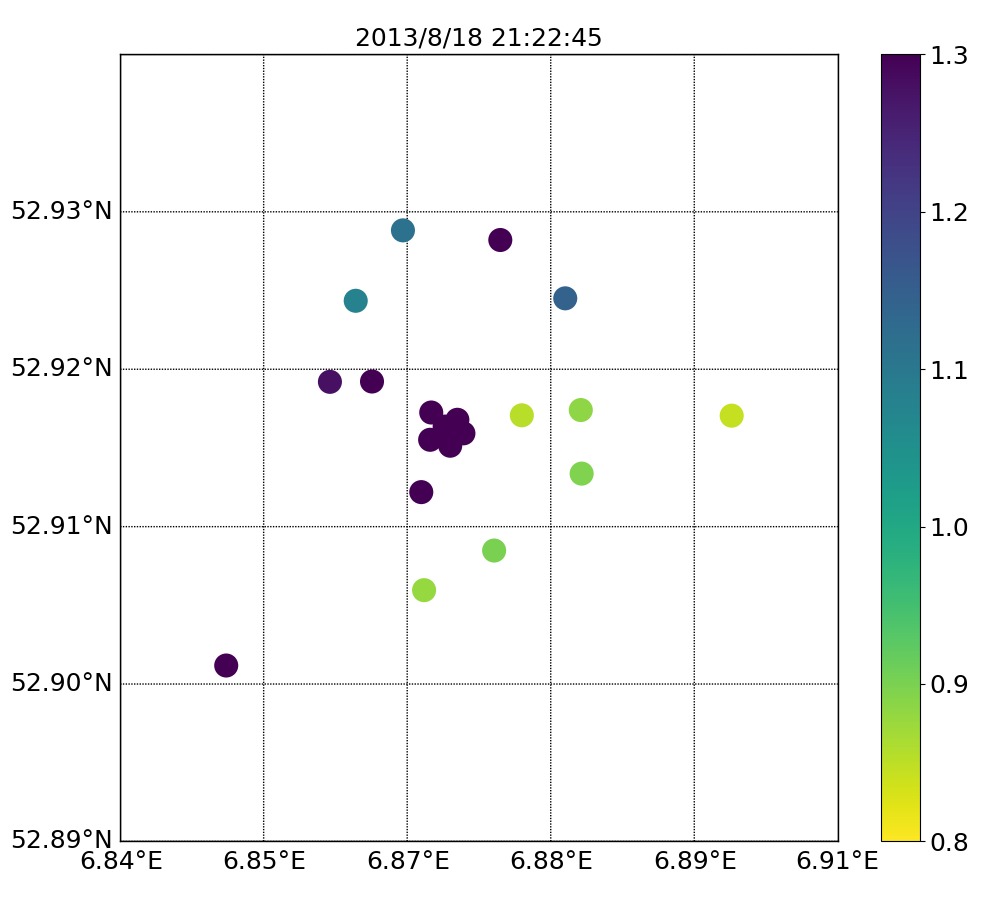}}
            \\
            \subfigure{\includegraphics[width=.266\linewidth]{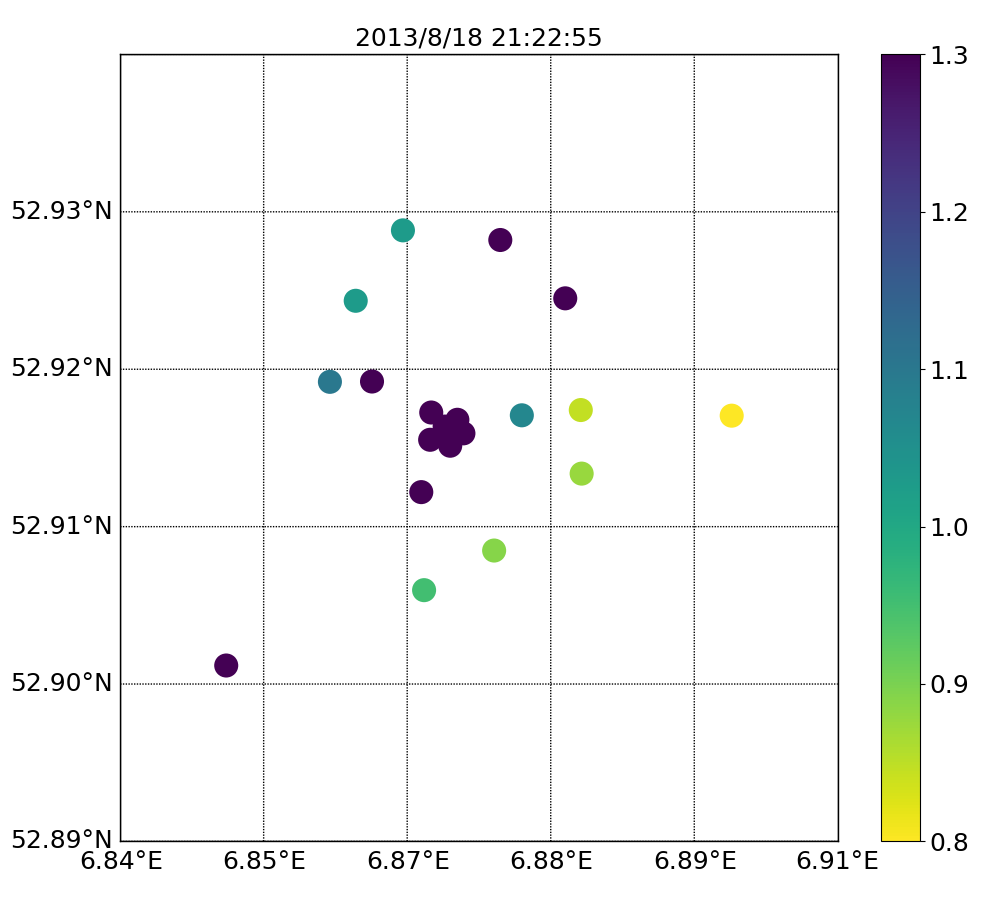}}
            \subfigure{\includegraphics[width=.266\linewidth]{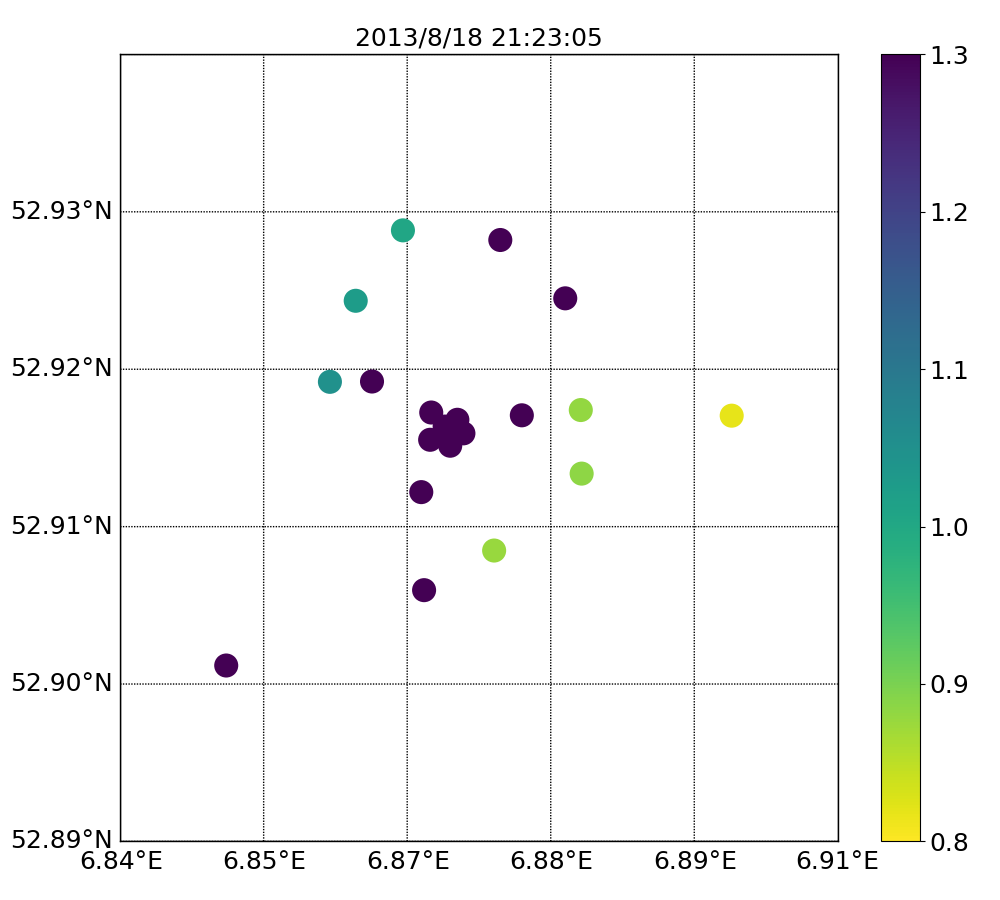}}
            \subfigure{\includegraphics[width=.266\linewidth]{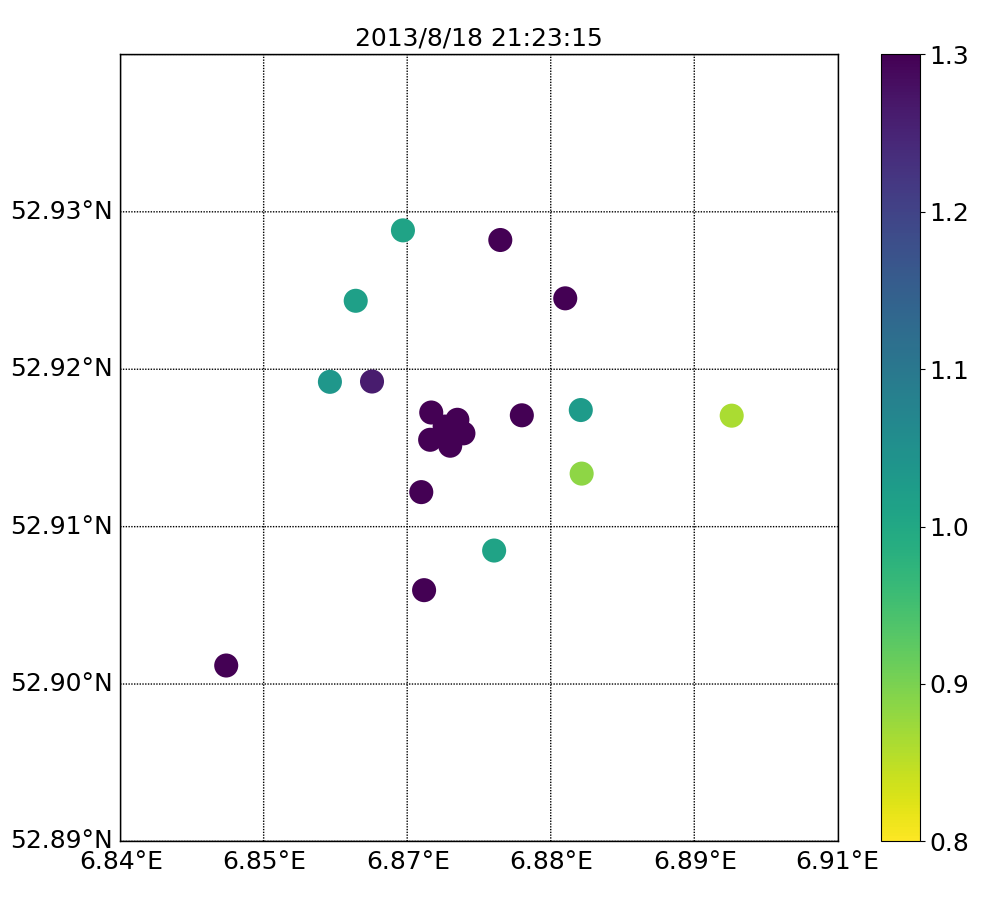}}
            \\
            \subfigure{\includegraphics[width=.266\linewidth]{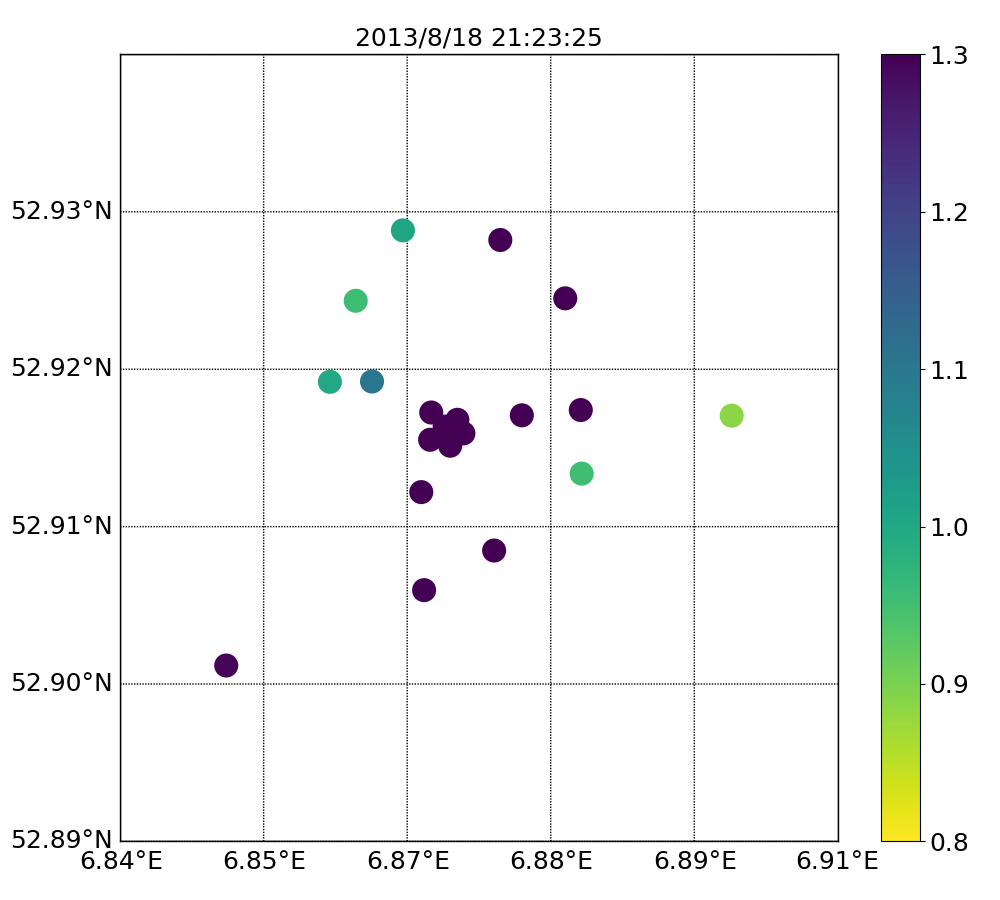}}
            \subfigure{\includegraphics[width=.266\linewidth]{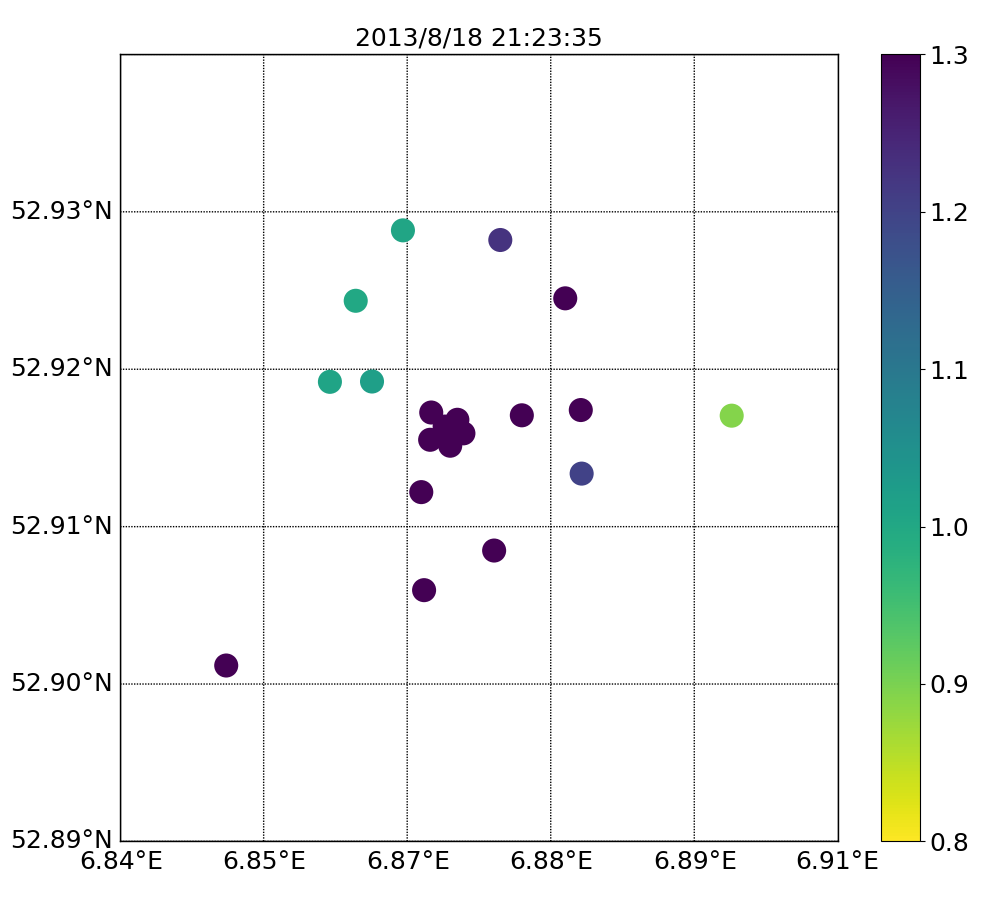}}
            \subfigure{\includegraphics[width=.266\linewidth]{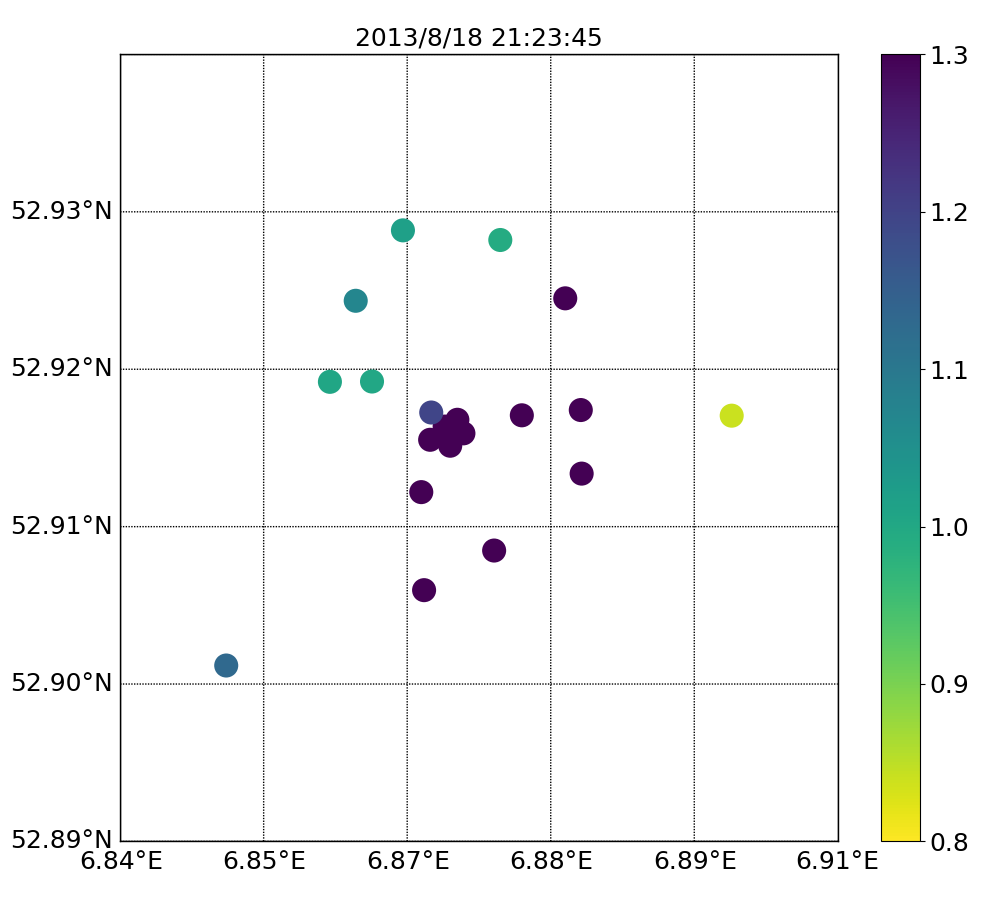}}
            \\
            \subfigure{\includegraphics[width=.266\linewidth]{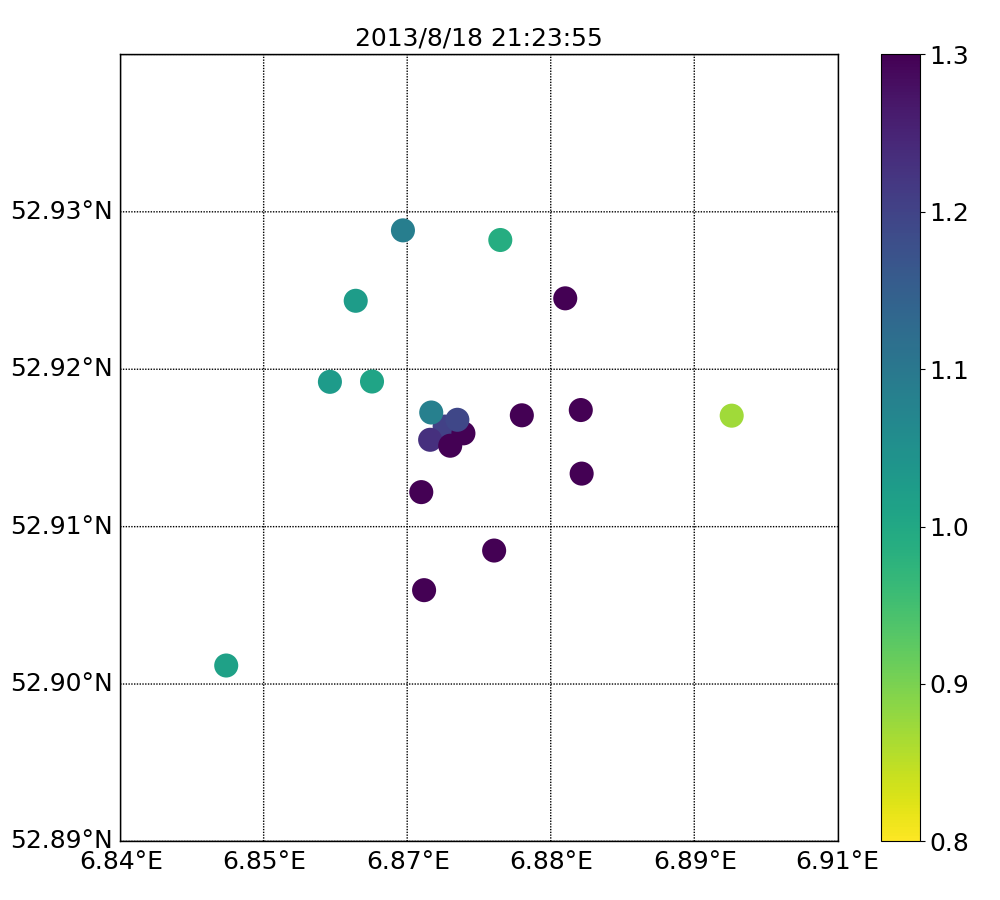}}
            \subfigure{\includegraphics[width=.266\linewidth]{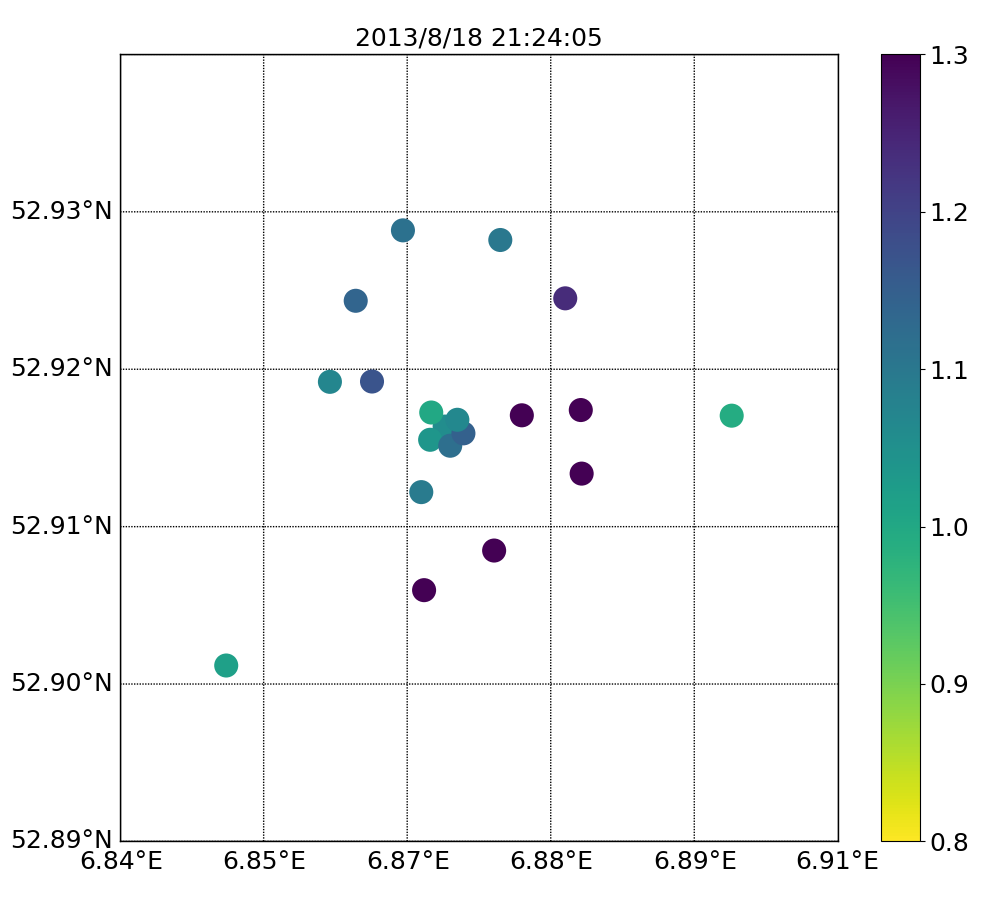}}
            \subfigure{\includegraphics[width=.266\linewidth]{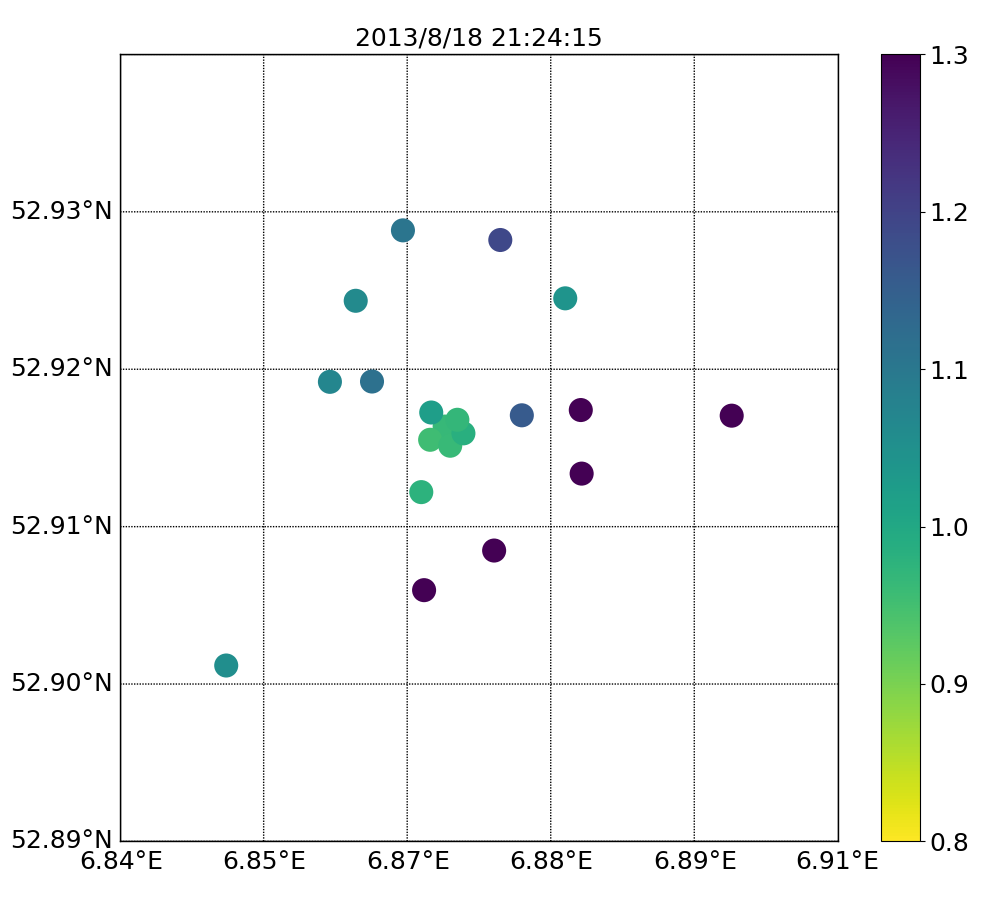}}
            \caption{Normalised intensities received by all core stations at an observing frequency of 44.13\,MHz, plotted on a geographical map of the stations.  The intensities are colour-coded using a colour scale from yellow to purple with a range of 0.8 to 1.3 respectively.  Times are at $\sim$10\,s intervals from 21:22:25\,UT at top left to 21:24:15\,UT at bottom right, and each plot uses data samples with an integration time of 0.92\,s.  Plot diameter is $\sim$4.5\,km.}
            \label{fig:scintillationmap}
        \end{figure}

        The core area of LOFAR contains 24 stations within an area with a diameter of $\sim$3.5\,km.  When viewing dynamic spectra from each of these stations it is clear that the scintillation pattern is mobile over the core (i.e., temporal shifts in the scintillation pattern are clear between stations) but does not necessarily evolve significantly.  Therefore, the flow of the scintillation pattern over the core stations may be viewed directly by simply plotting the intensity received, for a single subband, by each station on a map of geographical station locations, for data from successive time steps.  A movie (CasA\_20130818\_NL.mp4) of the scintillation pattern flow through the observation is published as an online supplement to this article.  The result, for 12 example time steps, is displayed in Figure \ref{fig:scintillationmap}, where a band of higher intensities can be seen to progress from north-west to south-east over the core.  It should be noted that the data were integrated in time to 0.92\,s for this purpose, to reduce both flicker due to noise and the duration of the movie.  This does not average over any scintillation structure in this observation; structure with periodicities shorter than one second would be obvious in the delay--Doppler spectra as an extension of the arc(s) to greater than 0.5\,Hz along the Doppler frequency axis.
        
\begin{figure}
            \centering
            \includegraphics[width=.95\linewidth]{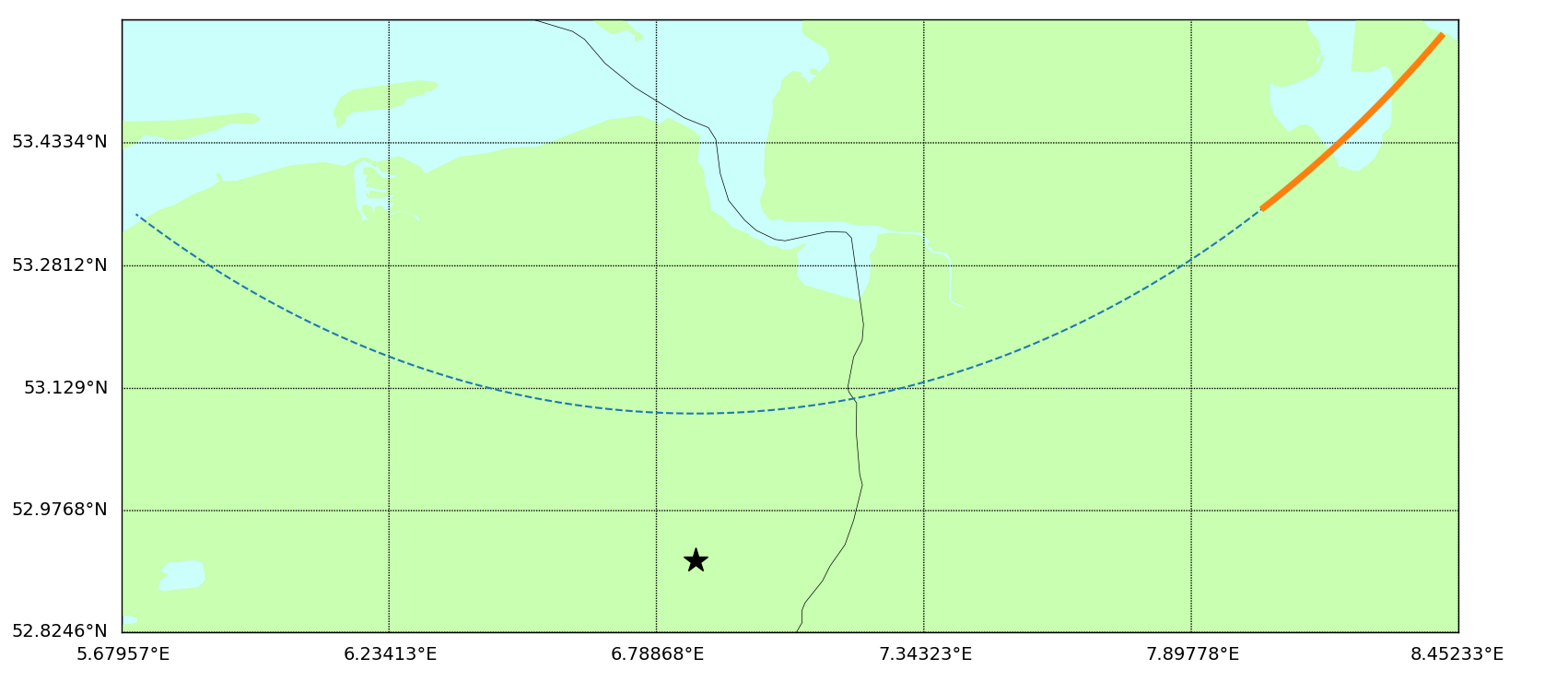}
            \caption{Map showing the track of the 200\,km pierce point of the line of sight from core station CS002 to Cassiopeia A from 2013-08-18T21:05:00 to 2013-08-19T04:05:00\,UT.  The thicker orange part of the track enhances the first hour of the observation.  The black line winding a path across the centre of the image is the location of the border between the Netherlands and Germany.  The location of CS002 is marked with a black star.}
            \label{fig:scintillationmaptrack}
        \end{figure}

        However, this is not the entire picture because the lines of sight from radio source to receivers are moving through the ionosphere as the Earth rotates, meaning that the scintillation pattern flow observed is a combination of flow due to the movement of density variations in the ionosphere and the movement of the lines of sight themselves through the ionosphere.  Since the speed with which any single point on a line of sight passing through the ionosphere is dependent on the altitude of that point (the so-called ionosphere ``pierce point''), this altitude needs to be either assumed or calculated to estimate a correction to the overall flow speed to obtain the natural ionospheric contribution.  This introduces a natural uncertainty into estimates of velocity.  Figure \ref{fig:scintillationmaptrack} shows the track of an ionospheric pierce-point at an assumed altitude of 200\,km (an altitude chosen as representative of a typical F-region altitude where large-scale plasma structures are commonly observed) for the line of sight from core station CS002 to the radio source Cassiopeia A through the 7-hour course of the observation to illustrate this movement.  Although not the subject of this paper, it is worth noting that an east to west flow seen later in the observation appears to be solely due to the lines of sight moving across a mostly static ionospheric structure (see the online movie), if the 200\,km pierce point is assumed, further illustrating the necessity to take accurate care of the contribution from line of sight movement when assessing ionospheric speeds.
        
        The movie of the scintillation pattern flow, assuming a 200\,km pierce point, shows a clear general north-west to south-east flow during the first hour of the observation, but also indicates some short (minutes) periods of confusion in which a north-east to south-west component might be just about discernable. Any second flow is likely to be associated with a second ionospheric layer and so warrants further investigation.
    
    \subsection{Estimating Velocities}
        
        The representation of the scintillation pattern flow in movie form gives a direct and broad picture of the flow pattern and is very helpful in discovering short time-scale changes in speed and direction.  However a cross-correlation analysis is still necessary to assess actual velocity(s).  Correlation functions are calculated as follows:-
        \begin{itemize}
            \item Time series' of intensity received by each station are calculated by averaging over the frequency band 55--65\,MHz, with these frequencies chosen as the scintillation pattern remains highly correlated over this band;
            \item For each three-minute data slice, advancing the start time of each successive slice by 10\,s:-
            \begin{itemize}
                \item Calculate auto- and cross- power spectra using intensities from every station pair within the LOFAR core;
                \item Apply low- and high-pass filters to exclude the DC-component and any slow system variation unlikely to be due to ionospheric effects, and white noise at the high spectral frequencies.  The white noise is also subtracted using an average of spectral power over the high frequencies;
                \item Inverse--FFT the power spectra back to the time domain to give auto- and cross-correlation functions.
            \end{itemize}
        \end{itemize}
        
\begin{figure}
            \centering
            \subfigure[]{\label{fig:velsd2105}\includegraphics[width=.45\linewidth]{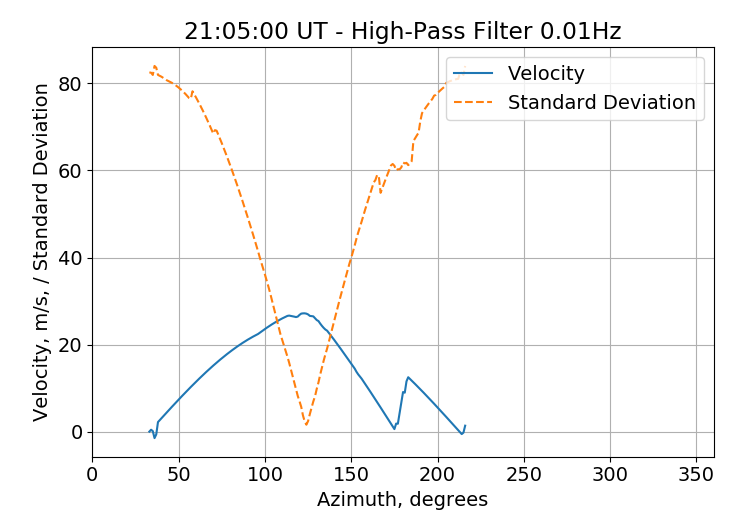}}
            \subfigure[]{\label{fig:velsd2115}\includegraphics[width=.45\linewidth]{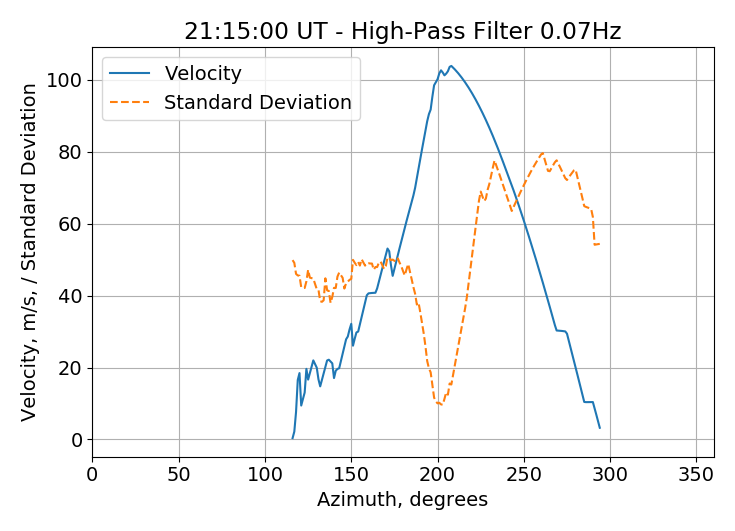}}
            \caption{Plots for single 3-minute time slices of the median velocity and standard deviation of velocities about the median versus azimuth direction, calculated from the range of velocities found from all cross-correlation functions with the baselines within each station pair re-calculated for each assumed azimuth direction, in the usual form, counting clockwise from north.  \subref{fig:velsd2105} Time slice commencing 21:05:00\,UT using cross-correlations calculated after applying a high-pass filter at 0.01\,Hz; \subref{fig:velsd2115} Time slice commencing 21:15:00\,UT using cross-correlations calculated after applying a high-pass filter at 0.07\,Hz. Note that the same y-axis is used for both velocity and standard deviation.}
            \label{fig:velocitysd}
        \end{figure}

        In the analysis the high- and low-pass filter values were set to 0.01\,Hz and 0.5\,Hz respectively.  This process results in a large set of cross-correlation functions for each time slice, each of which has an associated station-station baseline and a primary peak at, typically, a non-zero time delay from which a velocity can be calculated. 
        However, the direction of the scintillation pattern flow still needs to be found for calculation of the actual velocity.  For this, directions were assumed for each degree in the full 360--degree range of possible azimuth directions and the velocities re-calculated using the components of all baselines aligned with each assumed direction.  This results, for each time slice, in 360 sets of velocities and from each set a median velocity and standard deviation about the median can be calculated (the median is used as this is less susceptible to rogue data points than the mean).  The actual flow direction corresponds to the azimuth with the maximum median velocity and minimum standard deviation, as illustrated in Figure \ref{fig:velocitysd}.
        
        From this analysis the primary velocity of $\sim$20--40\,m\,s$^{-1}$ travelling from north-west to south-east is found, illustrated in Figure \ref{fig:velsd2105}, corresponding to the obvious scintillation pattern flow seen in the movie.  However, the presence of a second flow is still not obvious, although a hint of it can be seen in, for example, the second peak in the median velocity seen in Figure \ref{fig:velsd2105}.
        
\begin{figure}
            \centering
            \includegraphics[width=.7\linewidth]{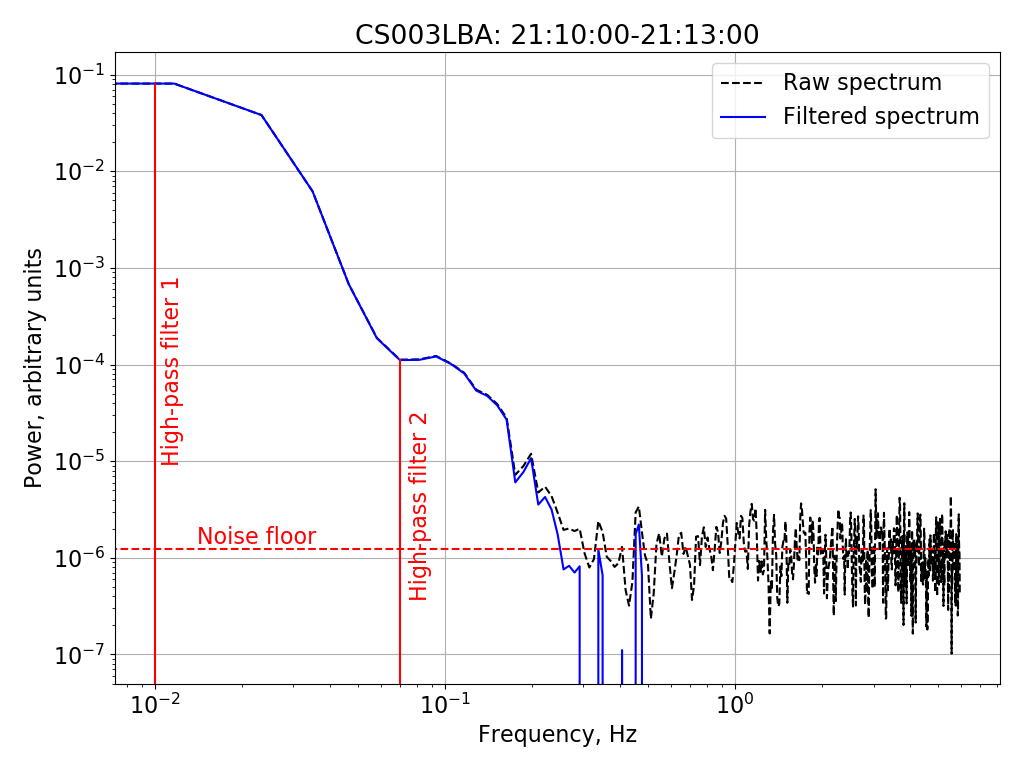}
            \caption{Example power spectrum calculated from three minutes of intensity data received by CS003.  The black curve is the raw spectrum, the blue curve is the filtered and noise-subtracted spectrum.  The locations of the low-pass filter and both high-pass filters used are illustrated.}
            \label{fig:powerspectrum}
        \end{figure}

        A closer look at the auto- power spectra yielded the key to finding the second flow.  Many spectra show a ``bump'' which can be viewed as being a second spectrum superposed on the main one.  This is illustrated in Figure \ref{fig:powerspectrum}.  To isolate this part of the spectrum, the spectra were re-filtered with a high-pass filter value of 0.07\,Hz (the low-pass filter value remained the same), and correlation functions re-calculated.  After following the same analysis as above to find median velocities and standard deviations, the second flow was found, as illustrated in Figure \ref{fig:velsd2115}.
        
\begin{figure}
            \centering
            \includegraphics[width=.95\linewidth]{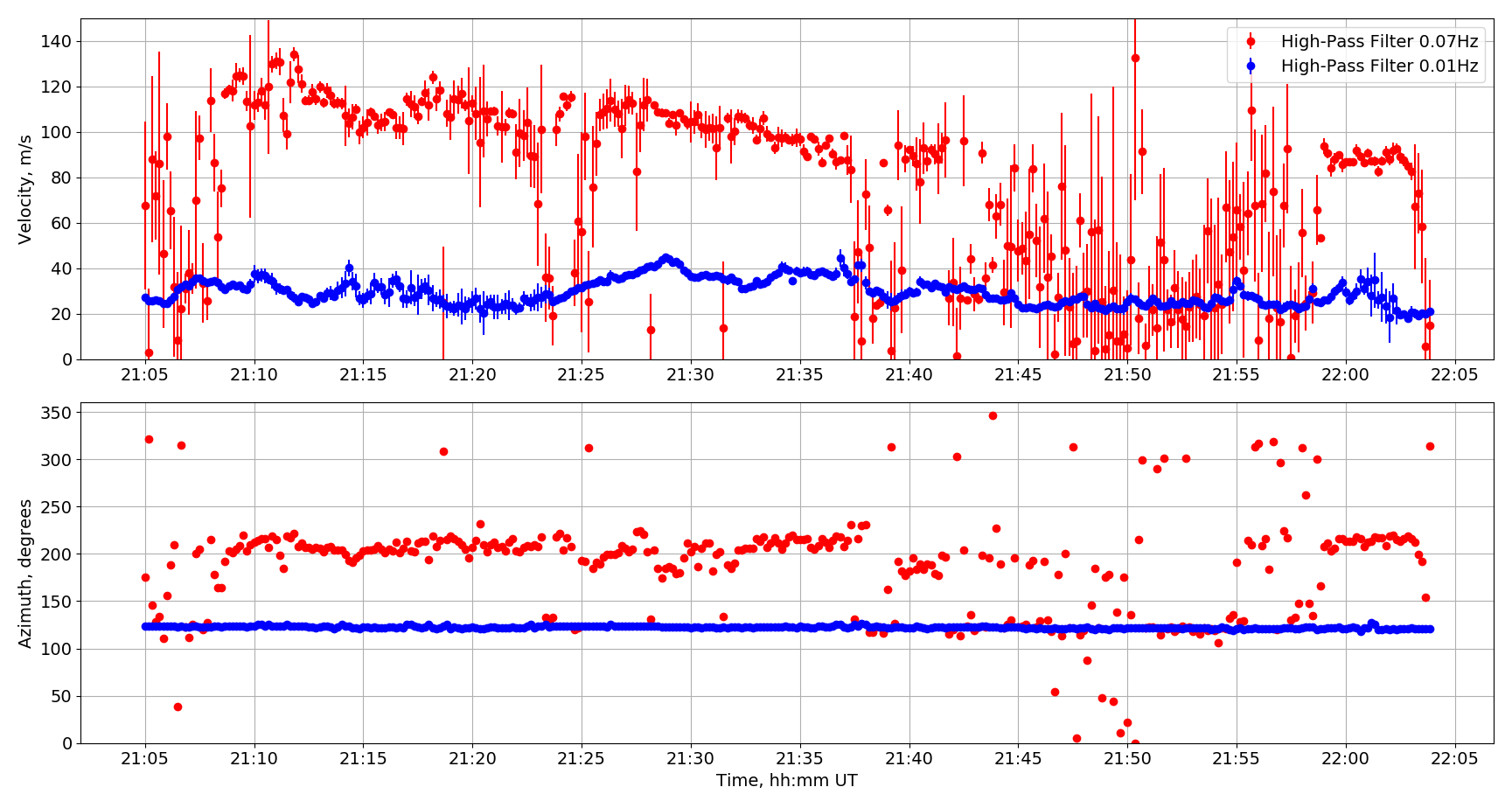}
            \caption{Top: Velocities calculated for the first hour of observation from cross-correlations created after filtering using the two different high-pass filter values.  Bottom: Directions of these velocities, in degrees azimuth.}
            \label{fig:velocities}
        \end{figure}

        The analysis, using both high-pass filter values, has been carried out for the full data set.  The velocities and associated directions in degrees azimuth for the first hour of the observation are given in Figure \ref{fig:velocities}.  Error bounds in the velocities are calculated as the standard deviation about the median of all velocity values available for the calculated azimuth direction.
        
        The higher velocity (henceforth labelled as the ``secondary velocity'') shows some scatter: Periods where the secondary velocity drops to around the primary velocity values are due to the secondary velocity not being detected at these times; in these cases, it can still be detected in short-duration drops of velocity if correlation functions are re-calculated using an even higher high-pass filter value (the bump in these spectra appears shifted to slightly higher spectral frequencies).  Values which decrease/increase towards/away from the primary velocity values likely represent a mix between the two velocities.  The larger error bars seen in velocities may also be indicative in some instances of the standard deviation being broadened by some velocity values being more dominated by the other flow.  The more extended period of scatter around 21:40 to 22:00\,UT is a period where the secondary velocity is less apparent and the secondary scintillation arc fades from the delay-Doppler spectra.  This indicates that the secondary structure is restricted in either space or time, either moving out of the field of view of the observation or ceasing for a period around 21:40\,UT.  It gives a first indication that the secondary velocity is associated with the secondary scintillation arc.
        
    \subsection{Estimating Scattering Altitudes}
        
        The velocities can now be used to estimate scattering altitudes, using the curvatures of the scintillation arcs and the simple formula given in Equation \ref{eqn:scintarc}.  Initially the movement of the line of sight through the ionosphere is not accounted for, since this correction also requires an estimate of the pierce-point altitude to be reasonably calculated.  Therefore an initial calculation of the scattering altitudes is made based on velocity values which are not corrected for this movement.
        
        Using the primary velocities and combining these with the curvatures of the primary arc (Figure \ref{fig:curves}) in Equation \ref{eqn:scintarc}, a range of distances, L, along the line of sight to the scattering region are found.  These distances are converted to altitudes by accounting for source elevation (Cas A increased in elevation from 55\,$^{\circ}$ to 64\,$^{\circ}$ during the first hour of observation).  This process resulted in a range of altitudes to the scattering region of 200 to 900\,km.  Doing the same for the secondary velocities and applying an arc curvature of 3.2$\pm$0.3 for the secondary scintillation arc gives estimated scattering altitudes of only $\sim$70\,km.  If the primary/secondary velocities are combined vice-versa with the secondary/primary arc curvatures respectively, then the resulting scattering altitudes are clearly unreasonable (the secondary arc, primary velocity combination gives estimated altitudes of only $\sim$10\,km for example), lending further credence to the secondary velocity being associated with the secondary arc.
        
        Velocity contributions from the line of sight movement are calculated as follows: For each time slice, t, the geographical locations beneath the pierce point of the line of sight through the ionosphere corresponding to the estimated scattering altitude at t are calculated, for both t and t + $\delta$t, where $\delta$t is taken as 3 minutes (the actual value is unimportant for this calculation).  A velocity and its direction are found from the horizontal distance between these two locations and the direction of travel from one to the other.  The general direction of the movement of the line of sight through the ionosphere is indicated by the orange line in Figure \ref{fig:scintillationmaptrack}.  Although the high scattering altitudes related to the primary scintillation arc and primary scintillation velocities lead to line-of-sight movements of up to $\sim$35\,m\,s$^{-1}$, this movement is almost perpendicular to the direction of the primary scintillation velocity, limiting the actual contribution to $\sim$ 5\,m\,s$^{-1}$.  The line of sight movement is, however, in a very similar direction to the secondary velocities but the low corresponding scattering altitudes also limit the contribution in this case to $\sim$5\,m\,s$^{-1}$.  
        
\begin{figure}
            \centering
            \includegraphics[width=0.95\linewidth]{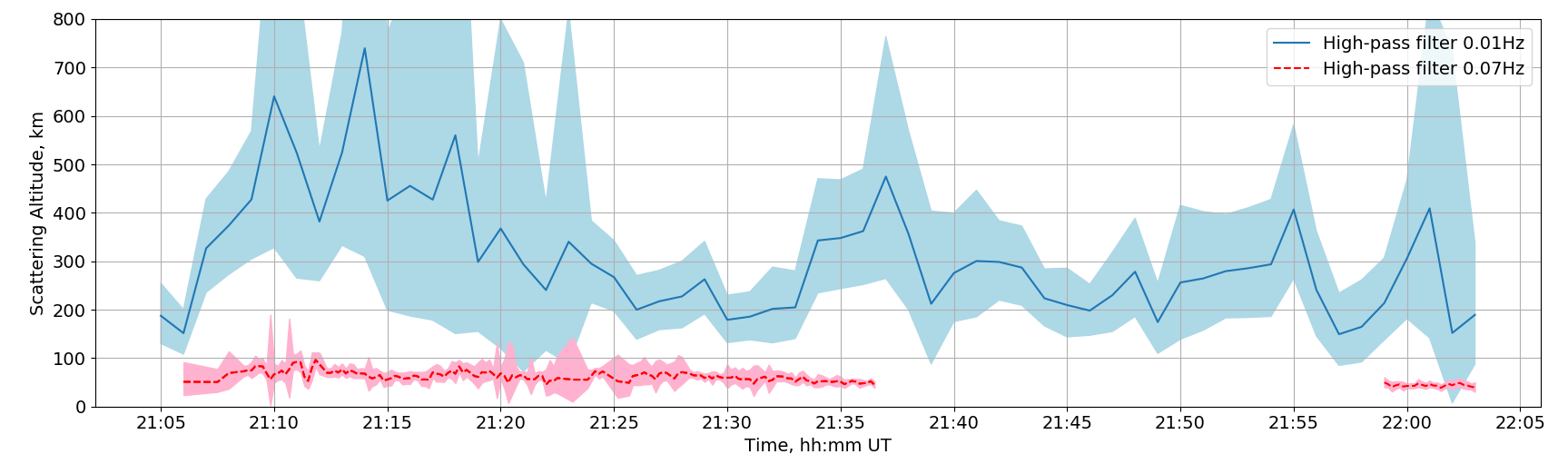}
            \caption{Scattering altitudes estimated using Equation \ref{eqn:scintarc}, the primary velocities and primary scintillation arc curvatures (blue curve) and the secondary velocities and the curvature of the secondary scintillation arc (red dashed curves).}
            \label{fig:scatalts}
        \end{figure}

        An iterative procedure is then followed to correct the scintillation velocities for line-of-sight movement at the calculated scattering altitudes, re-calculate these altitudes, and re-calculate the line-of-sight movement.  This procedure converges to a set of final scattering altitudes within 5 iterations.  These are presented in Figure \ref{fig:scatalts}, with error bounds taken as the lowest and highest possible altitudes resulting from applying this procedure using the lower and upper limits of the arc curvature and scintillation velocity error bounds.
        
        The range of scattering altitudes encompassed by the error bounds is quite large in some instances, particularly where the calculated altitudes are higher.  Although the square term for the velocity in Equation \ref{eqn:scintarc} could lead to the natural conclusion that the error in the velocity dominates the error in scattering altitude, the errors in the velocity calculations are, for the most part, relatively small.  Nevertheless, the error in the secondary velocity does appear to be the dominant error in the lower range of scattering altitudes (the red curves in Figure \ref{fig:scatalts}.  However, the dominant error for the higher range of scattering altitudes appears to be the scintillation arc curvatures, illustrating the importance of developing accurate fitting methods for these curvatures.  Despite these concerns, it is clear that scattering is seen from two layers in the ionosphere; the primary scintillation arc arises from scattering in the F-region and the secondary scintillation arc arises from scattering much lower down in the D-region.  Plasma decays by recombination with neutral species. In the F-region these densities are lower and so plasma lifetimes are longer than in the D-region. Typical plasma lifetimes in the F-region are of the order of hours, while they are of the order of minutes in the D-region. Hence the structures seen in each level may have a different source and time history.

\section{Conditions in the Ionosphere}

    We now investigate what the overall ionospheric conditions were at the time and hence the possible cause(s) of the scintillation seen by LOFAR at the time.
    
    \subsection{Geomagnetic Conditions}

\begin{figure}
            \centering
            \includegraphics[width=0.95\linewidth]{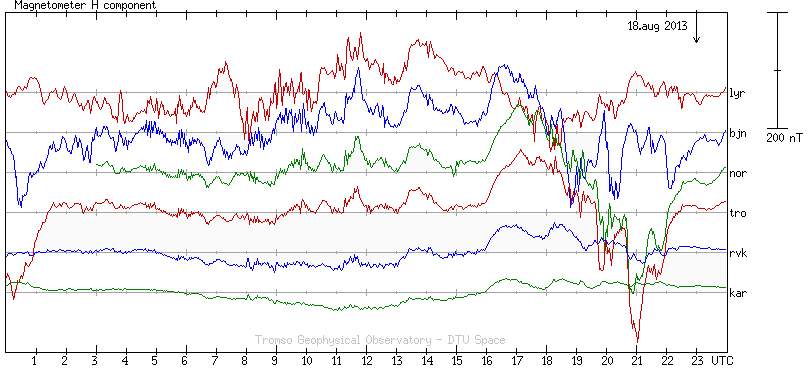}
            \caption{Traces of the H-component of the geomagnetic field recorded on 18 August 2013 by a selection of magnetometer stations from the Norwegian chain.  From top to bottom these are, along with their geomagnetic latitudes (2004, altitude 100\,km): Longyearbyen (75.31$^{\circ}$N), Bj\o rn\o ya (71.52$^{\circ}$N), Nordkapp (67.87$^{\circ}$N), Troms\o (66.69$^{\circ}$N), R\o rvik (62.28$^{\circ}$N), and Karm\o y (56.43$^{\circ}$N).}
            \label{fig:magnetometers}
        \end{figure}

        The overall geomagnetic conditions at the time are given in Figure \ref{fig:magnetometers}, which shows 24--hour traces of the H--component of magnetic field for a representative set of magnetometers from the Norwegian magnetometer chain for 18 August 2013.  Activity can be described as unsettled, with a minor substorm at high latitudes, peaking at the start of the LOFAR observation.  However, geomagnetic activity remains quiet further south, and Kp took a value of 1 at 21\,UT on 18th August 2013, indicating that this is unlikely to be a direct cause of the scintillation seen at LOFAR latitudes.  We therefore investigate whether TIDs were present at the time and whether these could be consistent with the scintillation seen by LOFAR.
        
    \subsection{Ionosonde Data}
    
\begin{figure}
            \centering
            \includegraphics[width=0.5\linewidth]{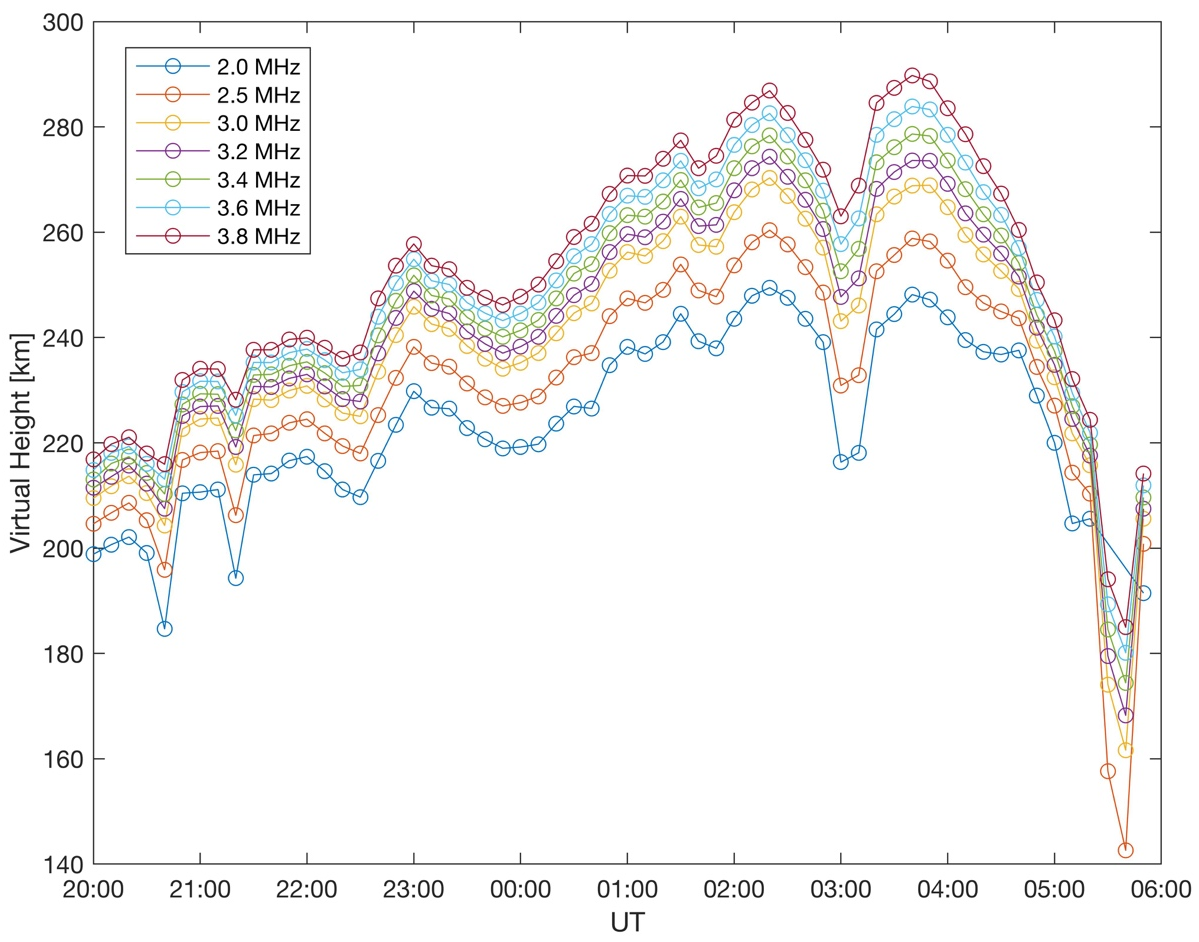}
            \caption{Multiple traces from the ionosonde in Chilton (UK) recorded between 20:00 18 August 2013 and 06:00 19 August 2013.}
            \label{fig:ionosonde}
        \end{figure}

        The presence of TIDs can be detected through the simultaneous appearance of wave-like structures on multiple sounding frequencies recorded by an ionosonde. This method is generally limited to a single point of observation and detection. The spatial extent of TIDs can be attempted by comparing multiple traces from different ionosondes, but this is limited by the low density of ionosondes in a given region.  Measurements from the ionosonde in Chilton (UK) do indeed suggest the presence of wave-like patterns which, in principle, could be due to a large-scale TID propagating southward and/or MSTID triggered by a local Atmospheric Gravity Wave (Figure \ref{fig:ionosonde}).
        
    \subsection{GNSS Data}
    
\begin{figure}
            \centering
            \includegraphics[width=0.5\linewidth]{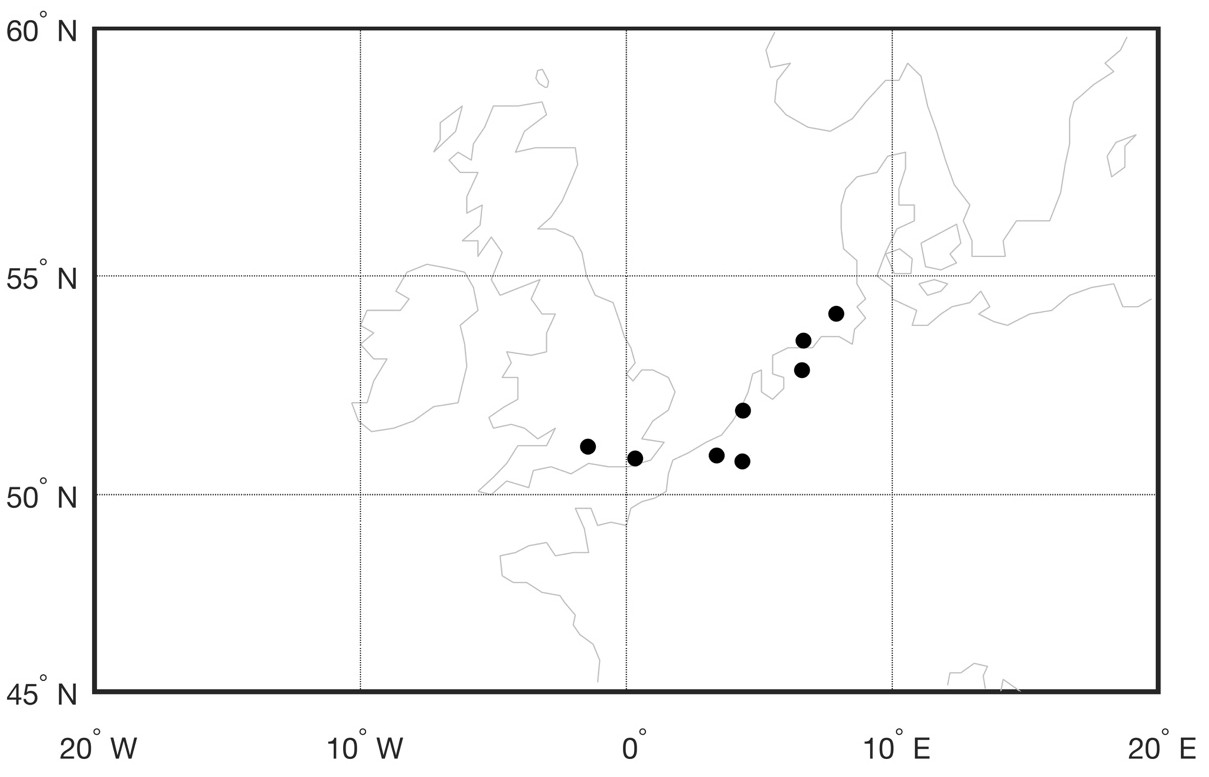}
            \caption{Map showing the locations of the GNSS stations used.}
            \label{fig:GPSstations}
        \end{figure}

        However, measurements from ground-based GNSS receivers offer a more comprehensive view of the characteristics of any MSTIDs present \citep{Kelley:2009}.  In the present study, we focus on perturbations in the slant Total Electron Content (STEC) observed over the evening of 18 August 2013 from a network of GNSS stations around the LOFAR core stations (see Figure \ref{fig:GPSstations}).  These stations are sufficient to infer the presence of TIDs and to infer the upper spatial scale-size limit of smaller-scale irregularities causing the intensity scintillation seen at LOFAR wavelengths.
        
        The presence of TID-induced perturbations can be deduced from the presence of wave-like residuals on the STEC calculated for each satellite-receiver pair.
        
        STEC was calculated and detrended following the methods of \citet{HernandezPajaresetal:2006}, with the detrending carried out according to:
        
        \begin{equation}
            \label{eqn:DSTECt}
            \Delta STEC \left( t \right) = STEC \left( t \right) - \frac{STEC \left( t+\tau \right) + STEC \left( t-\tau \right)}{2} \left[ TECu \right]
        \end{equation}
        where $\tau = 300s$.
        
        It is worth noting that the measured carrier phases $L_1$ and $L_2$ vary with time as a consequence of the motion of GNSS satellites relative to a given receiver on the Earth’s surface. As such, the spatial and temporal variabilities of ionisation gradients (such as those connected with TIDs and corresponding instabilities) become entangled. The various detrending methods (similar to equation \ref{eqn:DSTECt}) lead to an estimate of ionisation gradients by considering temporal gradients only, with spatial and temporal variabilities intrinsically entangled in the GNSS observations.
        
\begin{figure}
            \centering
            \subfigure[][]{\label{fig:satellite-c}\includegraphics[width=.5\linewidth]{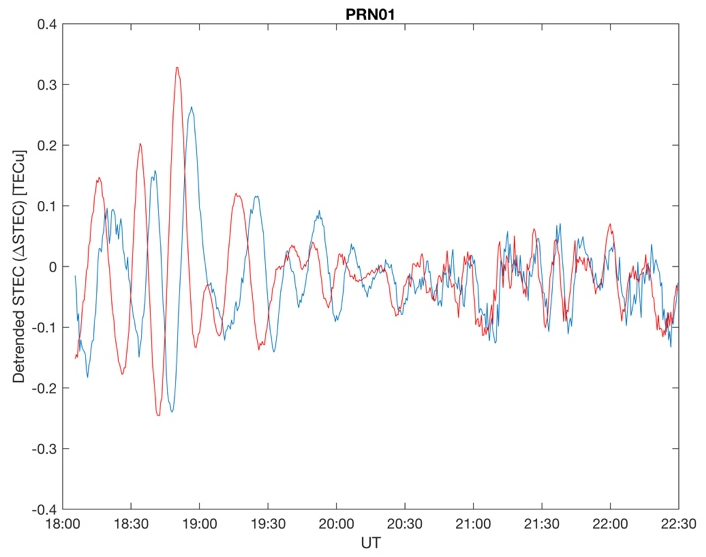}}
            \subfigure[][]{\label{fig:satellite-d}\includegraphics[width=.4\linewidth]{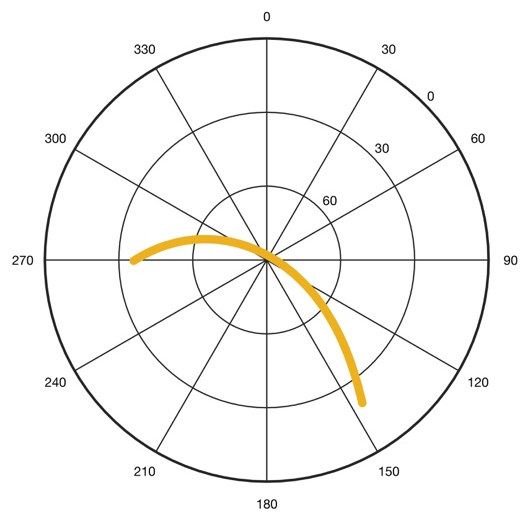}}
            \caption{Example of a satellite-station pair.  \subref{fig:satellite-c} PRN01 as observed on 18 August 2013 from Dentergem (DENT, blue line) and Bruxelles (BRUX, red line), both in Belgium, with baseline oriented from WNW to ESE; \subref{fig:satellite-d} azimuth/elevation plot for PRN01 as observed from Dentergem.}
            \label{fig:satellite}
        \end{figure}

        Figure \ref{fig:satellite} shows examples of wave-like residuals on STEC for one pair of GNSS stations (Dentergem and Bruxelles in Belgium) observing the same GNSS satellite.  The wave pattern is strongest over the first two hours shown (18:00 - 20:00\,UT) but then weakens considerably by the start of the LOFAR observation, although it remains evident.  STEC from the observations of both stations appears well--correlated, with the Bruxelles dataset lagging behind that of Dentergem.  Since Dentergem lies to the WNW of Bruxelles, this suggests a strong westerly component in the direction of travel, which could correspond with the secondary velocity seen by LOFAR.
        
\begin{figure}
            \centering
            \subfigure[]{\label{fig:STEC1819}\includegraphics[width=.45\linewidth]{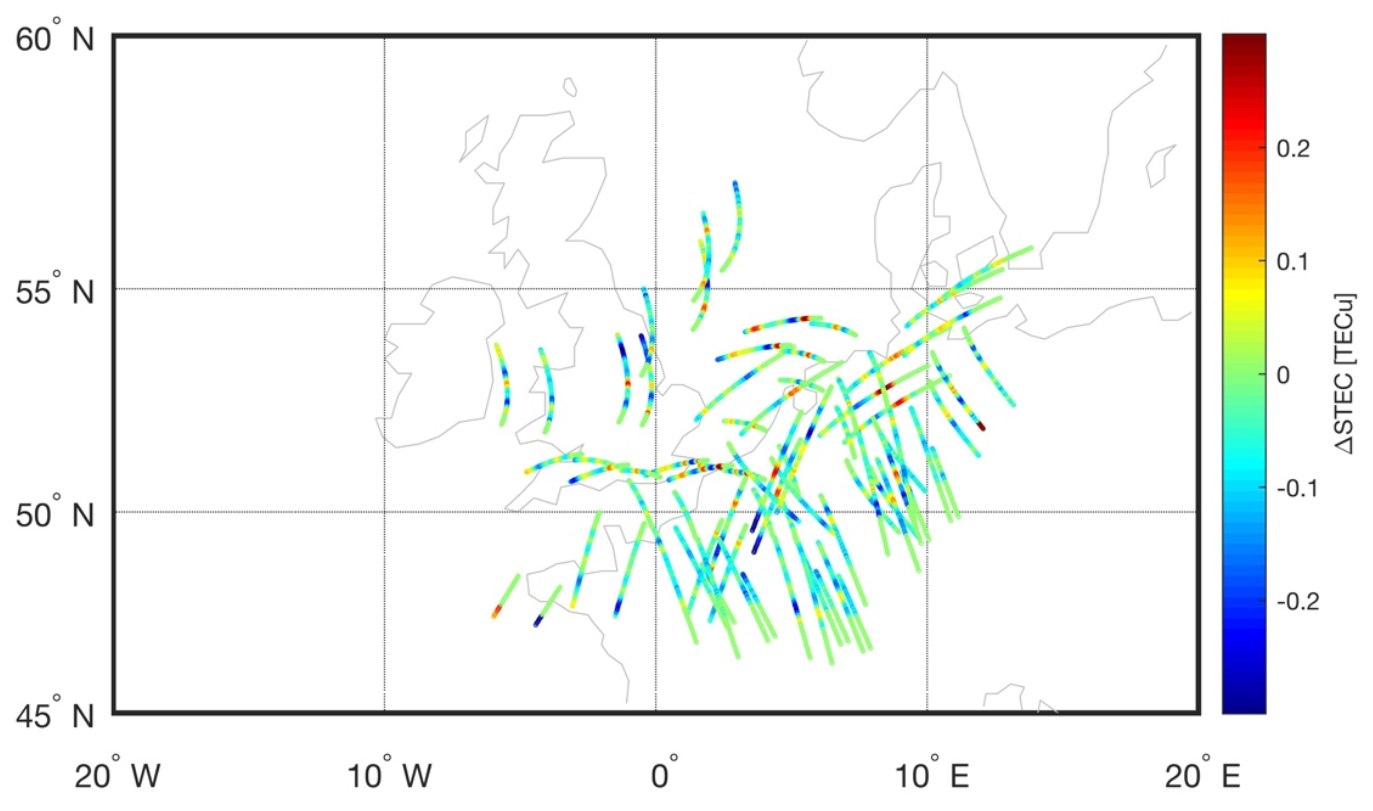}}
            \subfigure[]{\label{fig:STEC1920}\includegraphics[width=.45\linewidth]{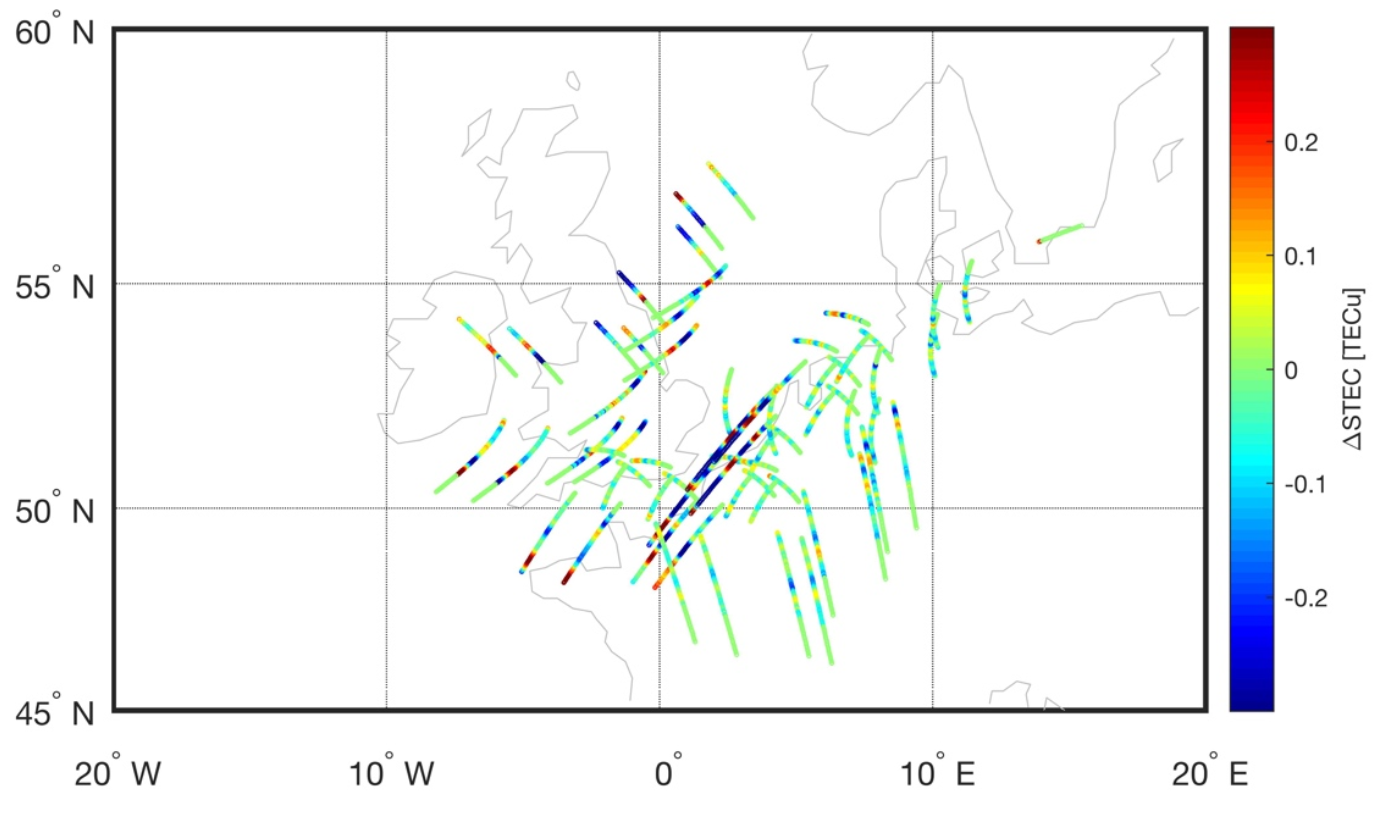}}
            \\
            \subfigure[]{\label{fig:STEC2021}\includegraphics[width=.45\linewidth]{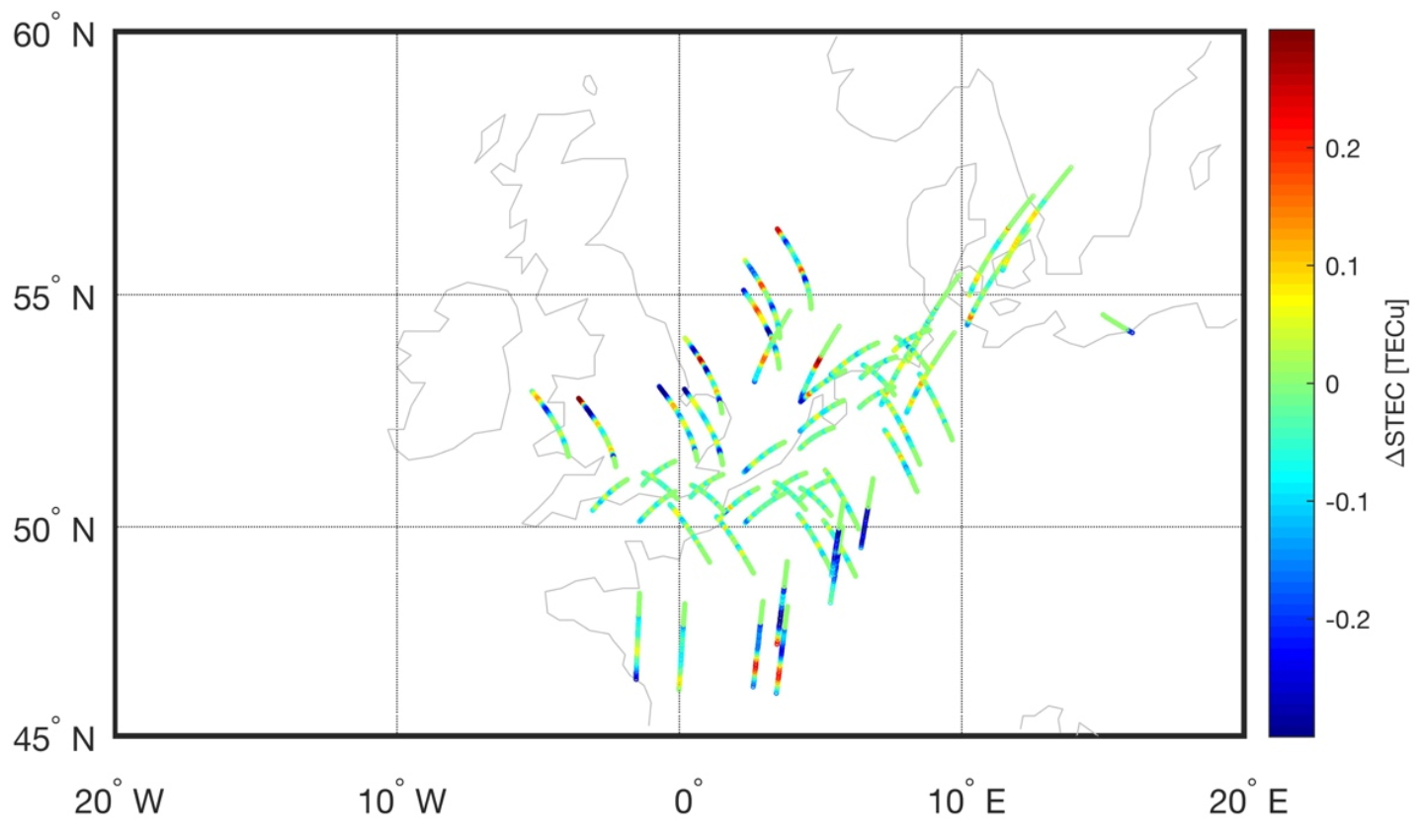}}
            \subfigure[]{\label{fig:STEC2122}\includegraphics[width=.45\linewidth]{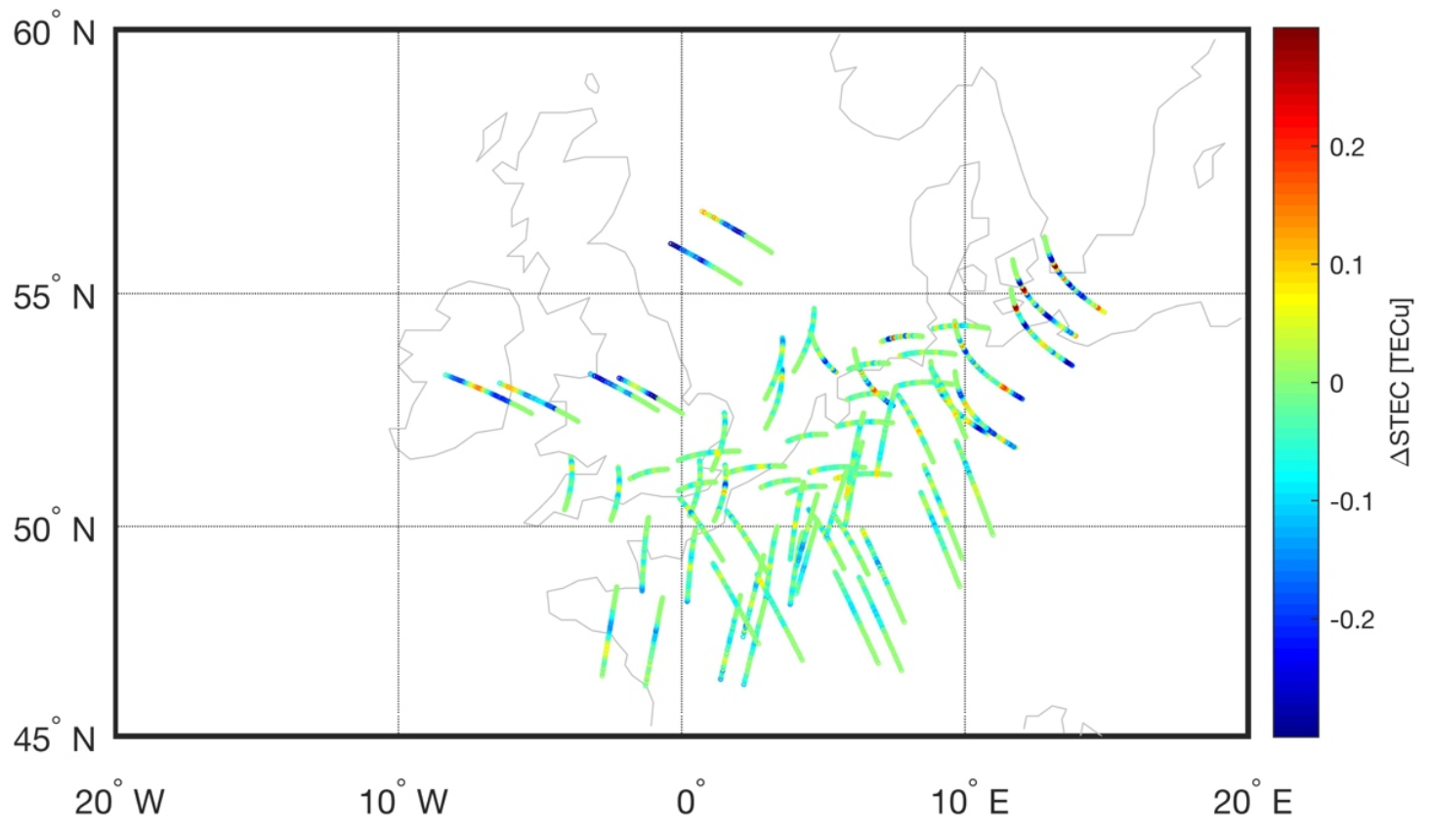}}
            \caption{Hourly geographical distribution of all STEC perturbations in the evening of 18 August 2013: \subref{fig:STEC1819} 18:00-19:00\,UT, \subref{fig:STEC1920} 19:00-20:00\,UT, \subref{fig:STEC2021} 20:00-21:00\,UT, and \subref{fig:STEC2122} 21:00-22:00\,UT.}
            \label{fig:STEC}
        \end{figure}

        Figure \ref{fig:STEC} shows hourly plots of the overall geographical distribution of the STEC residuals calculated for all satellite passes seen within each hour by the GNSS stations used.  The patterns shown in Figure \ref{fig:STEC} suggest a spatially and temporally varying propagation of MSTID wavefronts with components along the NE-SW as well as the NW-SE directions. Furthermore, the examples shown in Figure \ref{fig:STEC} also indicate the presence of smaller-scale ionisation structures in proximity to the wavefronts of the MSTIDs. This suggests that the scintillation seen by LOFAR is likely associated with the perpendicular propagation of two MSTIDs.  However, the STEC variations here are also seen to fade by the start of the LOFAR observation.
        
\begin{figure}
            \centering
            \subfigure[]{\label{fig:PSDTEC2021}\includegraphics[width=.47\linewidth]{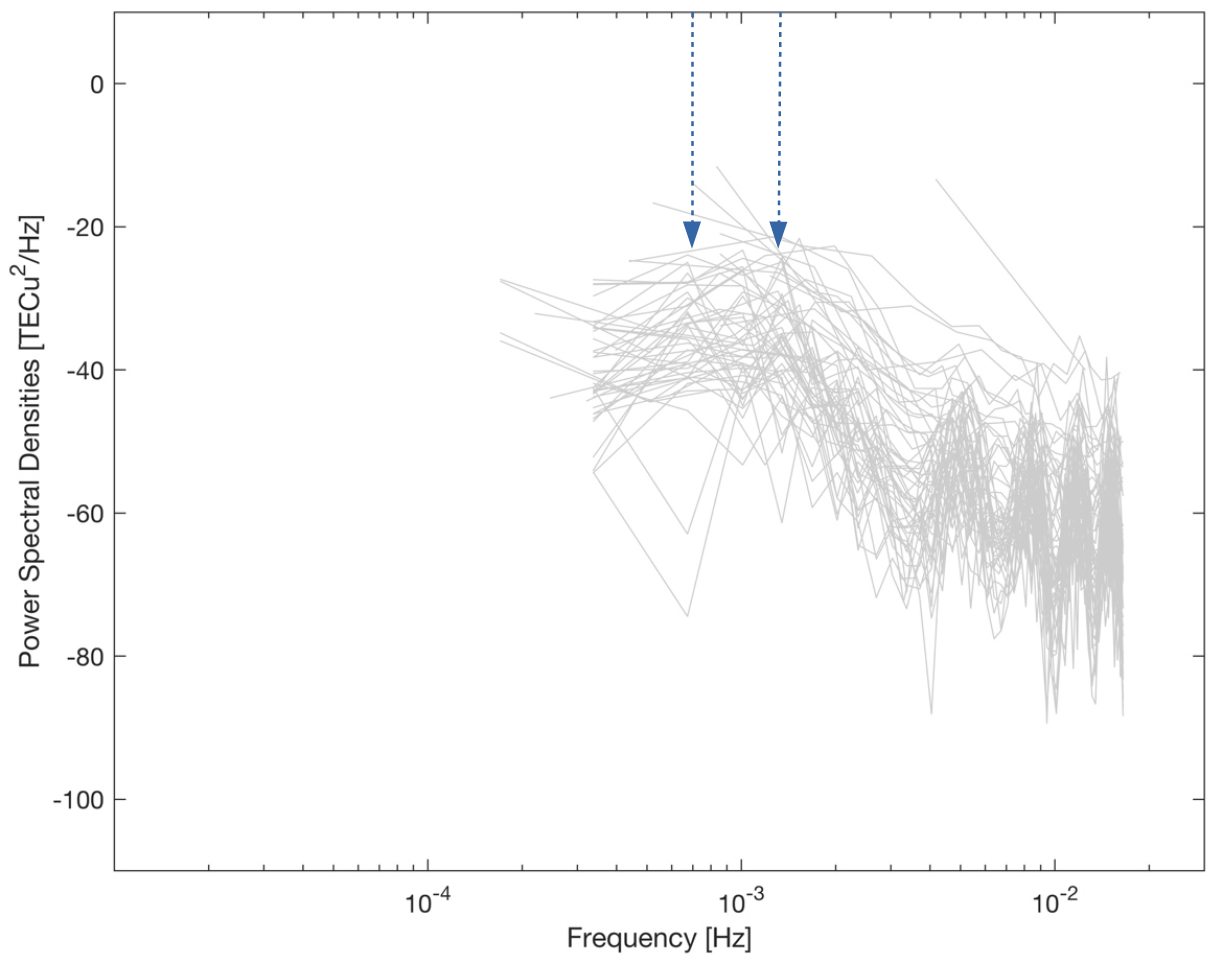}}
            \subfigure[]{\label{fig:PSDTEC2122}\includegraphics[width=.47\linewidth]{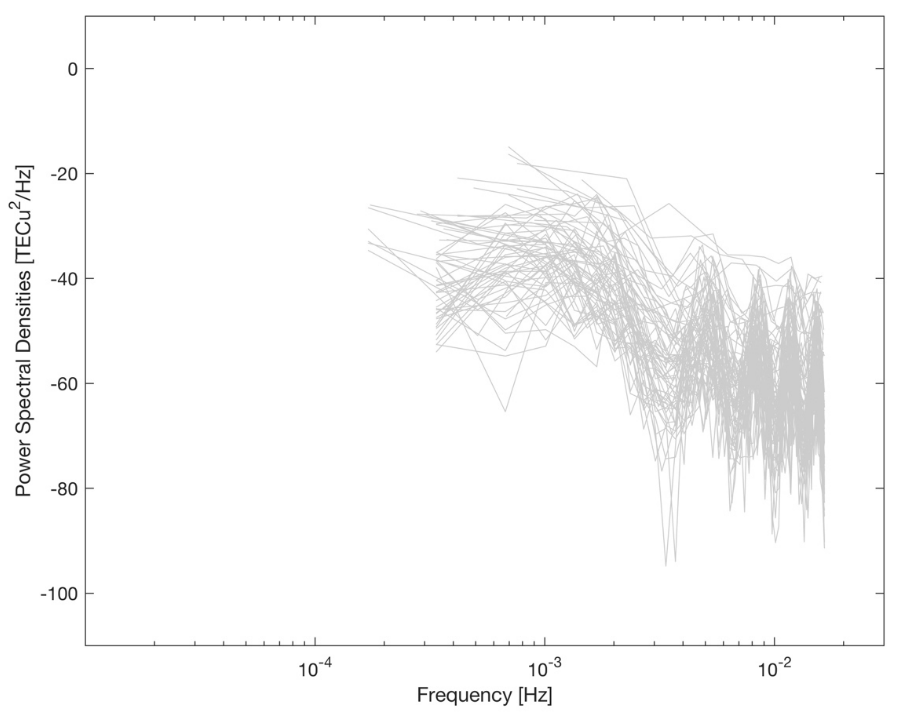}}
            \caption{Power Spectral Densities of all the TEC residuals considered during the hours \subref{fig:PSDTEC2021} 20:00-21:00\,UT and \subref{fig:PSDTEC2122} 21:00-22:00\,UT.  The arrows indicate the two components considered in the text.}
            \label{fig:PSDTEC}
        \end{figure}

        A further illustration looks at the overall power spectral densities for the STEC residuals on all satellite--receiver pairs considered here over the hourly periods 20:00\,UT to 21:00]\,UT and 21:00\,UT to 22:00\,UT (Figure \ref{fig:PSDTEC}).  The earlier hour is chosen alongside the hour covering the LOFAR data period as this better displays the components seen in the spectra  The temporal frequencies f can be converted into spatial scales L by assuming a given velocity V$_{REL}$ for the motion of the ionospheric structures across a GNSS raypath. That is:
        
        \begin{equation}
            \label{eqn:L}
            L = \frac{V_{REL}}{f}
        \end{equation}
        where V$_{REL}$ = V$_{IONO}$-V$_{SAT}$ is the relative velocity between the velocity of the ionospheric structures and the scan velocity of a single raypath (at the same shell height). V$_{SAT}$ can be of the order of a few tens of m\,s$^{-1}$ at 300\,km.
        
        There appear to be two main components in the energy cascade from larger to smaller ionisation scales: one with a period of $\sim$1666\,s, and another component with a period of $\sim$666\,s.  Taking V$_{REL}$ to be $\sim$100\,m\,s$^{-1}$ (the secondary velocity seen by LOFAR as this is in a south-westerly direction and the example GNSS data in Figure \ref{fig:satellite} indicate a westerly component), these periodicities  correspond to spatial scales of the order of 166\,km and 66\,km respectively. Beyond these scales the STEC analysis is limited by the sensitivity of the technique \citep{Tsugawaetal:2007}, as the Power Spectral Densities reach the noise floor (Figure \ref{fig:PSDTEC}). These orders of magnitudes suggest the presence of a larger--scale TID together with a smaller--scale TID \citep{Kelley:2009}, while the energy cascade that can be observed through the Power Spectral Densities indicates that the large--scale structure breaks down into small--scale structures, likely owing to some instability mechanism. 
        
    
    \subsection{Estimation of Scale Sizes of Plasma Structures}
        
        The scale sizes of the plasma structures causing the scintillation seen by LOFAR can also be calculated.  The variations in the intensity of the received signal are caused by irregularities with a spatial scale size ranging from the Fresnel dimension to an order of magnitude below this value \citep{Basuetal:1998}.  The Fresnel length D$_{F}$ is related to the wavelength of the radio wave $\lambda$ and the line of sight distance from the receiver to the scattering region L:
        
        \begin{equation}
            \label{eqn:DF}
            D_{F} = \sqrt{2 \lambda L}
        \end{equation}
        
        The Fresnel length was calculated for plasma structures at altitudes of 70\,km, 200\,km, 350\,km and 700\,km, elevations of 55$^{\circ}$ and 64$^{\circ}$, and at frequencies of 25.19\,MHz, 35.15\,MHz and 60.15\,MHz, and the results are shown in Table \ref{tab:Fresnel}. The altitudes were chosen to cover the range of altitudes identified for the primary and secondary features in the LOFAR analysis, with the addition of 350\,km as this altitude is commonly used within studies using GNSS satellites. The elevations of the radio source at the start and the end of the first hour of observation were used to establish the range of Fresnel scales for each altitude. The frequencies were chosen to match Figure \ref{fig:timespectra}. 
        
        Table \ref{tab:Fresnel} shows that the Fresnel length ranges between $\sim$1\,km and $\sim$5\,km and therefore the plasma structures causing the variations in signal intensity are likely to have a spatial scale size between $\sim$100\,m and $\sim$5\,km.  The velocities calculated from the LOFAR data indicate that such structures would take tens of seconds to pass through the source-to-receiver line and the intensity variations in the observed signal occur on a similar timescale.
        
        \begin{table}[]
            \centering
            \begin{tabular}{c|c|c|c|c}
                \hline
                 Altitude & 70\,km & 200\,km & 350\,km & 700\,km \\
                 Frequency & & & & \\
                 \hline
                 25.19\,MHz & 1.4 & 2.3--2.4 & 3.0--3.2 & 4.3--4.5 \\
                 35.15\,MHz & 1.2 & 1.9--2.0 & 2.6--2.7 & 3.6--3.8 \\
                 60.15\,MHz & 0.9 & 1.5--1.6 & 2.0--2.1 & 2.8--2.9 
            \end{tabular}
            \caption{The Fresnel length at altitudes of 70\,km, 200\,km, 350\,km and 700\,km for three different frequencies received by LOFAR station CS002.  The ranges represent calculation using the source elevation for the start and for the end of the first hour of observation.  Values are in km.}
            \label{tab:Fresnel}
        \end{table}
        
\section{Further Discussion}
    
    Geomagnetic activity was low in the mid-latitudes at the time, so enhanced activity was unlikely to be the direct cause of the scintillation observed.  However, a weak sub-storm was seen at high latitudes and this reached its peak at the time of the start of the observation.  An analysis of GNSS and ionosonde data reveals the presence of an MSTID travelling in the north-west to south-east direction.  The larger-scale nature of this TID, and its direction of travel, are strongly consistent with the primary velocity and F-region scattering altitudes seen in the LOFAR observation.  It is possible that this TID was caused by the geomagnetic activity at high latitude, but this is not confirmed.  Simultaneously, an MSTID is also present travelling in a north-east to south-west direction which would most likely be associated with an atmospheric gravity wave propagating up from the neutral atmosphere.  The smaller--scale nature of it, its direction of travel, and likely low-altitude source make it highly consistent with the secondary velocity and D--region scattering altitudes observed by LOFAR.  
    
    The amplitude of TID activity observed through GNSS STEC residuals decreased after 20:00\,UT (as visible from Figure \ref{fig:satellite} as well as from the comparison of hourly geographical maps in Figure \ref{fig:STEC}).  However, the LOFAR observation did not start until 21:05\,UT and the presence of scintillation on the radio frequencies observed by LOFAR remained significant for much of the first hour of observation.  Whilst the presence of MSTIDs seems evident from the ionosonde multiple traces and GNSS STEC residuals in the region considered, their signatures do not appear simultaneously above the LOFAR core stations between 21:00\,UT and 22:00\,UT.  This can be explained by the inability of GNSS to detect smaller amplitudes in STEC residuals, as the noise floor is encountered for observations with pierce points above the core LOFAR stations (Figures \ref{fig:STEC} and \ref{fig:PSDTEC}).  The scale sizes of plasma structures calculated for the LOFAR data indicate that these are an order of magnitude lower than those estimated from GNSS STEC.  Smaller ionisation scales developing, for example, through the Perkins instability could induce scintillation on the VHF radio frequencies received by LOFAR but not on the L-band frequencies of GNSS.  Hence, scintillation from these mid-latitude smaller-scale ionisation structures, formed through the Perkins instability in conjuction with the presence of TIDs, is likely to be what is detected through LOFAR.
    

\section{Conclusions and Outlook}

    This paper presents the results from one of the first observations of ionospheric scintillation taken using LOFAR, of the strong natural radio source Cassiopeia A taken overnight on 18--19 August 2013.  The observation exhibited moderately strong scattering effects in dynamic spectra of intensity received across an observing bandwidth of 10--80\,MHz.  Delay--Doppler spectra from the first hour of observation showed two discrete parabolic arcs, one with a steep and variable curvature and the other with a shallow and static curvature, indicating that the scintillation was the result of scattering through two distinct layers in the ionosphere.  
    
    A cross-correlation analysis of the data received by stations in the LOFAR core reveals two different velocities in the scintillation pattern: A primary velocity of $\sim$20-40\,m\,s$^{-1}$ is observed travelling in a north-west to south-east direction, which is associated with the primary parabolic arc and altitudes of the scattering layer varying in the range $\sim$200--700\,km.  A secondary velocity of $\sim$110\,m\,s$^{-1}$ is observed travelling in a north-east to south-west direction, which is associated with the secondary arc and a much lower scattering altitude of $\sim$60--70\,km.  The latter velocity is associated with a secondary ``bump'' seen at higher spectral frequencies in power spectra calculated from time series' of intensities, indicating that it is more strongly associated with smaller--scale structure in the ionosphere.   
    
    GNSS and ionosonde data from the time suggest the presence of two MSTIDs travelling in perpendicular directions.  The F-region scattering altitudes calculated from the LOFAR primary scintillation arc and primary velocity, and the larger density scales associated with this, suggest that this is associated with a larger--scale TID seen in GNSS data potentially resulting from high--latitude geomagnetic activity.  The D-region scattering altitudes of the secondary arc and secondary velocity suggest an atmospheric gravity wave source for a smaller-scale TID.  These TIDs trigger an instability which leads to the breakdown of the large-scale density structure into smaller scales, giving rise to the scintillation observed.  In the mid-latitude ionosphere the Perkins mechanism is the most likely instability and the features of the smaller-scale density variations observed seem consistent with this.  To the best of our knowledge this is the first time that two TIDs have been directly observed simultaneously at different altitudes.
    
    This observation demonstrates that LOFAR can be a highly valuable tool for observing ionospheric scintillation in the mid--latitudes over Europe and enables methods of analysis to be used which give greater insight into the likely sources of scattering and could be used to improve modelling of them.  With a far greater range of frequencies (multi--octave if the LOFAR high--band is also used) and fine sampling both across the frequency band and in time, LOFAR observations offer a wider sensitivity than that available to GNSS measurements.  The analysis techniques shown in this paper also demonstrate that LOFAR can observe ionospheric structures at different altitudes simultaneously; a capability not commonly available for GNSS observations.  It also complements these measurements by probing potentially different scintillation regimes to those observed by GNSS.
    
    Since this observation was taken, many more have been carried out under a number of projects, recording ionospheric scintillation data at times when the telescope would otherwise be idle.  These demonstrate a wide range of scintillation conditions over LOFAR, some of which are seen only very occasionally and perhaps by only one or two of the international stations, illustrating the value to be had by monitoring the ionosphere at these frequencies.  A Design Study, LOFAR4SpaceWeather (LOFAR4SW -- funded from the European Community’s Horizon 2020 Programme H2020 INFRADEV-2017-1 under grant agreement 777442) currently underway will design a possible upgrade to LOFAR to enable, amongst other space weather observations, ionospheric monitoring in parallel with the regular radio astronomy observations.  Such a design, if implemented, would enable a full statistical study of ionospheric scintillation at these frequencies, alongside the advances in scintillation modelling and our understanding of the ionospheric conditions causing it which can be gleaned in focussed studies such as that presented here.

\begin{acknowledgements}
    This paper is based on data obtained with the International LOFAR Telescope (ILT) under project code ``IPS''. LOFAR \citep{vanHaarlemetal:2013} is the Low Frequency Array designed and constructed by ASTRON. It has observing, data processing, and data storage facilities in several countries, that are owned by various parties (each with their own funding sources), and that are collectively operated by the ILT foundation under a joint scientific policy. The ILT resources have benefitted from the following recent major funding sources: CNRS-INSU, Observatoire de Paris and Université d'Orléans, France; BMBF, MIWF-NRW, MPG, Germany; Science Foundation Ireland (SFI), Department of Business, Enterprise and Innovation (DBEI), Ireland; NWO, The Netherlands; The Science and Technology Facilities Council, UK; Ministry of Science and Higher Education, Poland.  The work carried out at the University of Bath was supported by the Natural Environment Research Council [Grant Number NE/R009082/1] and by the European Space Agency/Thales Alenia Space Italy [H2020-MOM-TASI-016-00002].  We thank Troms\o~ Geophysical Observatory, UiT the Arctic University of Norway, for providing the lyr, bjn, nor, tro, rvk, and kar magnetometer data.  The Kp index and the Chilton ionosonde data were obtained from the U.K. Solar System Data Centre at the Rutherford Appleton Laboratory.  Part of the research leading to these results has received funding from the European Community's Horizon 2020 Programme H2020-INFRADEV-2017-1under grant agreement 777442.
\end{acknowledgements}


\bibliography{LOFAR_ionosphere_20130818}
   

\end{document}